\documentclass{aa}

\usepackage{graphicx}
\usepackage{txfonts}

\begin{document}

\title{Anatomy of a buckling galactic bar}

\author{Ewa L. {\L}okas
%\inst{1}
}

\institute{Nicolaus Copernicus Astronomical Center, Polish Academy of Sciences,
Bartycka 18, 00-716 Warsaw, Poland\\
\email{lokas@camk.edu.pl}}

%\date{Received September 15, 1996; accepted March 16, 1997}

\abstract{
Using $N$-body simulations we study the buckling instability in a galactic bar forming in a Milky Way-like galaxy. The
galaxy is initially composed of an axisymmetric, exponential stellar disk embedded in a spherical dark matter halo.
The parameters of the model are chosen so that the galaxy is mildly unstable to bar formation and the evolution is
followed for 10 Gyr. A strong bar forms slowly over the first few gigayears and buckles after 4.5 Gyr from the start of the
simulation becoming much weaker and developing a pronounced boxy/peanut shape. We measure the properties of the bar at
the time of buckling in terms of the mean acceleration, velocity, and distortion in the vertical
direction. The maps of these quantities in face-on projections reveal characteristic quadrupole patterns which wind up
over a short timescale. We also detect a secondary buckling event lasting much longer and occurring only in the outer
part of the bar. We then study the orbital structure of the bar in periods before and after the first buckling. We find
that most of the buckling orbits originate from x1 orbits supporting the bar. During buckling the ratio of the vertical
to horizontal frequency of the stellar orbits decreases dramatically and after buckling the orbits obey a very tight
relation between the vertical and circular frequency: $3 \nu = 4 \Omega$. We propose that buckling is initiated by the
vertical resonance of the x1 orbits creating the initial distortion of the bar that later evolves as kinematic bending
waves.}


\maketitle

\section{Introduction}

We have known for a few decades now that galactic disks are inherently unstable and form bars easily
\citep{Miller1970, Ostriker1973}. Although the presence of a dark matter halo modifies the process
\citep{Athanassoula2003}, this instability has been accepted as the main channel of bar formation in galaxies
\citep[for a review see][]{Athanassoula2013}. Another possible scenario for the origin of bars involves tidal
interactions of galactic disks with perturbers of different sizes \citep{Noguchi1996, Miwa1998, Lokas2014, Lokas2016,
Gajda2017, Gajda2018, Lokas2018, Peschken2019}. Such tidally induced bars result from sufficiently strong
tidal deformations and are in all aspects similar to those formed in isolation, although perhaps have smaller
pattern speeds.

Independently of their origin, strong bars tend to undergo an event of buckling instability during their evolution
\citep{Combes1981, Pfenniger1991, Raha1991, Athanassoula2016, Smirnov2019, Lokas2019}. The phenomenon has been studied
using $N$-body simulations and shown to involve significant distortions of the bar out of the disk plane that thicken
and weaken the bar, but do not destroy it; although the presence of gas has been argued to suppress buckling
\citep{Debattista2006, Berentzen2007, Villa2010}. The instability may occur more than once in the lifetime of a bar,
with the second episode usually lasting longer and happening in the outer parts of the bar \citep{Martinez2006}. The
event leaves behind a distinct boxy/peanut shape in the inner part of the bar similar to bulges of some late-type
galaxies \citep{Debattista2004, Athanassoula2005, Bureau2006, Yoshino2015, Erwin2016, Erwin2017, Li2017,
Savchenko2017}. Buckling probably also took place in the bar of the Milky Way, as evidenced by the boxy/peanut
structure \citep{Weiland1994, Ciambur2017}, possibly leaving its traces in the presently observable phase-space
\citep{Khoperskov2019}.

The nature of buckling instability remains unclear. \citet{Combes1990} and \citet{Pfenniger1991} were the first to
propose that the instability results from trapping x1 orbits of the bar at vertical resonances. However, such
resonances are expected to apply only to banana-like orbits that have been shown to contribute only a little to the
final boxy/peanut shape \citep{Patsis2002, Portail2015, Valluri2016, Abbott2017, Patsis2018}. An alternative hypothesis
for the origin of the instability involves the ratio of the vertical to horizontal velocity dispersion of the stars in
the bar and relates it to the fire-hose instability \citep{Toomre1966, Raha1991, Merritt1991, Merritt1994}. In this
picture, the instability is supposed to be triggered by a sufficiently low dispersion ratio of the order of 0.3,
characteristic of strong bars.

\begin{figure}
\centering
\includegraphics[width=9cm]{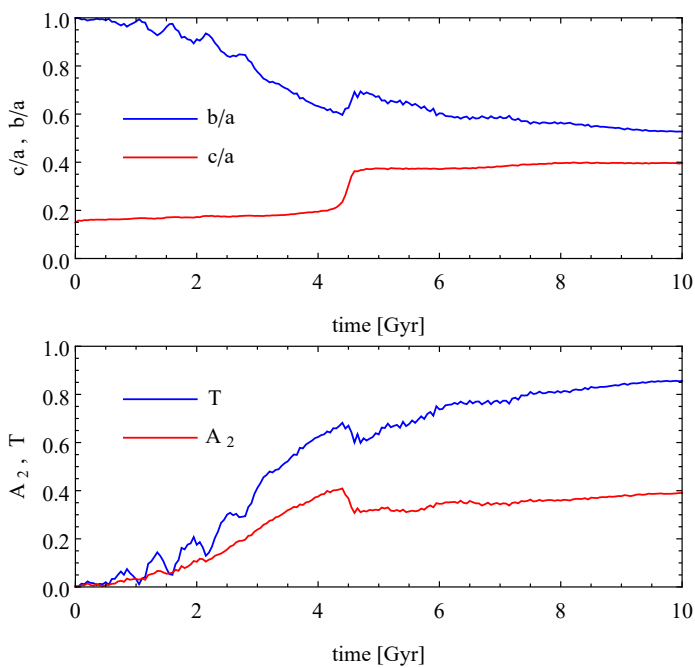}
\caption{Evolution of the shape of the stellar component in time. The upper panel shows
the evolution of the axis ratios $b/a$ (intermediate to longest axis) and $c/a$ (shortest to longest axis). The lower
panel shows the triaxiality parameter $T = [1-(b/a)^2]/[1-(c/a)^2]$ and the bar mode $A_2$. Measurements were made for
stars within the radius of $2 R_{\rm d}$.}
\label{shape}
\end{figure}

In \citet{Lokas2019} we used simulations of tidally induced galactic bars from \citet{Lokas2018} to study buckling
instability. Such configurations have the advantage that the same initial galaxy model can be used to create bars of
different strength by different tidal forcings. We demonstrated that for such bars there is no direct relation between
the ratio of the vertical to radial velocity dispersion and the susceptibility of the bar to buckling, suggesting that
buckling is due to vertical orbital resonances rather than the fire-hose instability. This conclusion was supported by an
approximate calculation of the vertical and horizontal resonances, which were shown to coincide during buckling.
Although tidal interactions may be an important or even dominant scenario for the formation of galactic bars
\citep{Peschken2019}, the difficulty in using tidally induced bars is that during tidal interactions in addition to bars
strong spiral arms are formed which later wind up and may disturb the orbital structure.

In this work we therefore revisit the issue of buckling instability using a simulation of an isolated galaxy
initially composed of an axisymmetric exponential disk and a dark matter halo. The bar develops in the disk over a few
gigayears of evolution and then buckles. In addition to different measures of distortion introduced in \citet{Lokas2019} we
characterize buckling by studying the orbital structure of the bar before and after buckling. For this purpose we apply
the spectral analysis of stellar orbits as pioneered by \citet{Binney1982}. This approach has been developed over the
years and successfully used to study the orbital structure of different potentials, including galactic bars
\citep{Miralda1989, Papaphilippou1996, Carpintero1998, Valluri1998, Merritt1999, Voglis2007, Ceverino2007, Deibel2011,
Valluri2010, Valluri2012, Valluri2016, Portail2015}, although most of these studies relied on test particle orbits
evolved in static external potentials. \citet{Gajda2016} demonstrated that the orbital structure can be reliably
measured also `in vivo', that is in a live bar evolving in an $N$-body simulation, even in configurations that are not
stationary by construction, for example in dwarf galaxies orbiting the Milky Way.

The paper is organized as follows. In Section 2 we provide the details of the simulation used in this study and
characterize the evolution of the bar formed in the galaxy. Section 3 presents the description of the buckling
instability in terms of the mean acceleration, streaming velocity, and distortion of the stars out of the disk
plane in the vertical direction, measured both radially and in the face-on projection. The orbital structure of the
bar before and after buckling is studied in Section 4 and the discussion follows in Section 5.

\begin{figure}
\centering
\includegraphics[width=7.7cm]{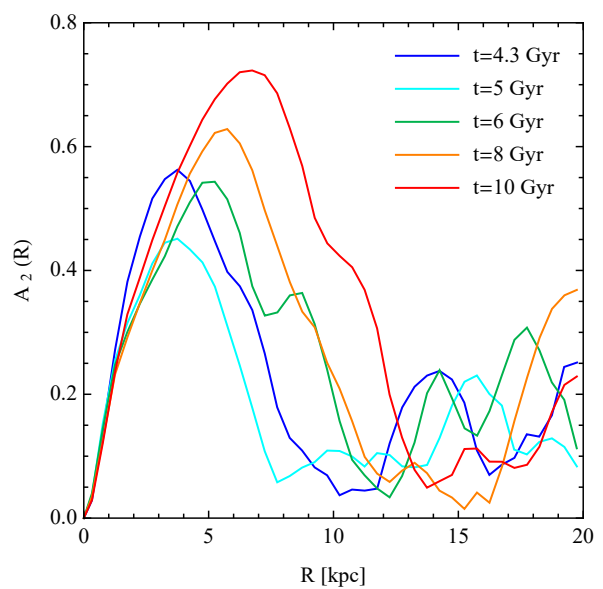}
\caption{Profiles of the bar mode $A_2 (R)$ at different times.}
\label{a2profiles}
\end{figure}

\begin{figure}
\centering
\hspace{0.7cm} \includegraphics[width=3.5cm]{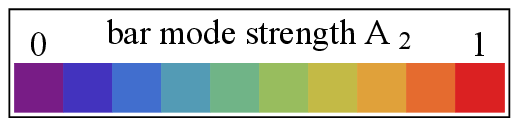}
\includegraphics[width=8.9cm]{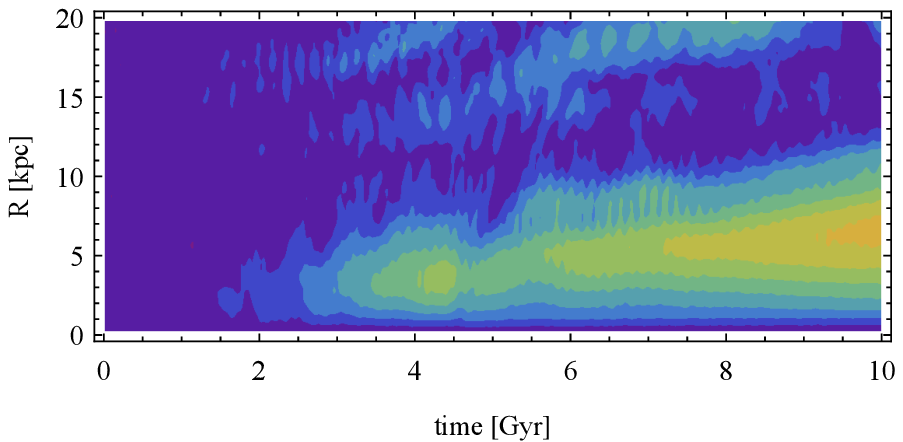}
\caption{Evolution of the profiles of the bar mode $A_2 (R)$ over time.}
\label{a2modestime}
\end{figure}

\section{The simulated bar}

For the purpose of this study we ran an $N$-body simulation of a two-component galaxy with properties similar to the
Milky Way. The exact parameters were chosen, after a few trial simulations, so that the galaxy forms the bar relatively
slowly, with it becoming strong enough to buckle after a few gigayears of evolution.

The galaxy initially contained two components: a spherical dark matter halo and an exponential disk. The structural
parameters of the galaxy were the following: its dark matter halo had an Navarro-Frenk-White \citep{Navarro1997}
profile with a virial mass $M_{\rm H} = 10^{12}$ M$_{\odot}$ and concentration $c=25$ while the exponential disk had a
mass $M_{\rm D} = 4.5 \times 10^{10}$ M$_{\odot}$, scale-length $R_{\rm D} = 3$ kpc, and thickness $z_{\rm D} = 0.42$
kpc. The central value of the radial velocity dispersion was $\sigma_{R,0}=120$ km s$^{-1}$. The minimum value of the
Toomre parameter for this model at $2.5 R_{\rm D}$ was $Q=1.73$. We note that with this choice of parameters the
contribution to the rotation curve from the disk and the halo in the inner parts is very similar.

The $N$-body realization of the galaxy was initialized using the procedures described in \citet{Widrow2005} and
\citet{Widrow2008} with each component containing $10^6$ particles. The evolution of the galaxy was followed for 10 Gyr
with the GIZMO code \citep{Hopkins2015}, an extension of the widely used GADGET-2 \citep{Springel2001, Springel2005},
saving outputs every 0.05 Gyr. The adopted softening scales were $\epsilon_{\rm D} = 0.03$ kpc and $\epsilon_{\rm H} =
0.06$ kpc for the disk and halo of the galaxy, respectively, within the ranges recommended by \citet{Hopkins2018}.

\begin{figure}
\centering
\includegraphics[width=9cm]{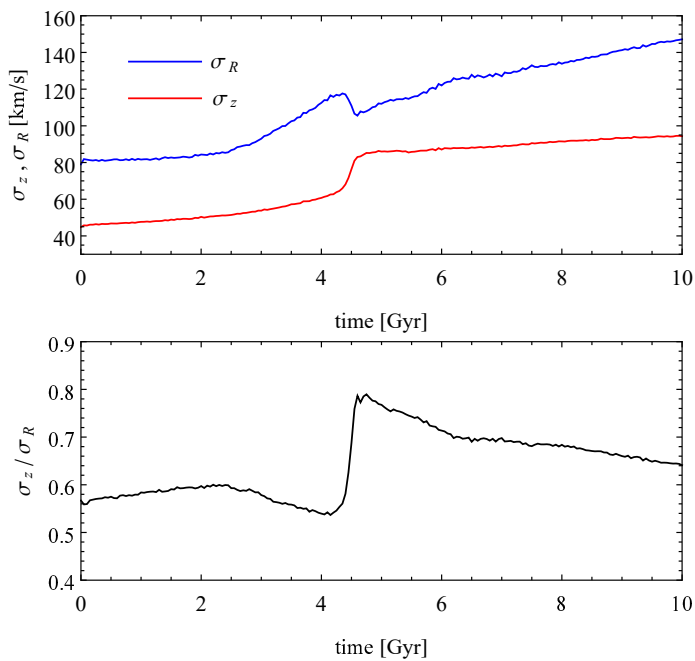}
\caption{Evolution of velocity dispersions of the stellar component over time. The upper panel shows the velocity
dispersions along the cylindrical radius $\sigma_R$ and along the vertical direction
$\sigma_z$ while the lower panel plots the ratio $\sigma_z/\sigma_R$. Measurements were made for
stars within the radius of $2 R_{\rm D}$.}
\label{kinematics}
\end{figure}

\begin{figure}
\centering
\hspace{0.7cm} \includegraphics[width=6cm]{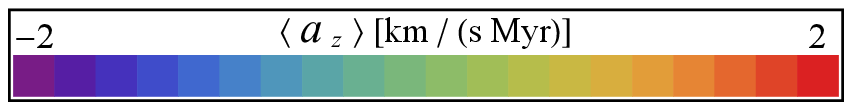}
\includegraphics[width=8.9cm]{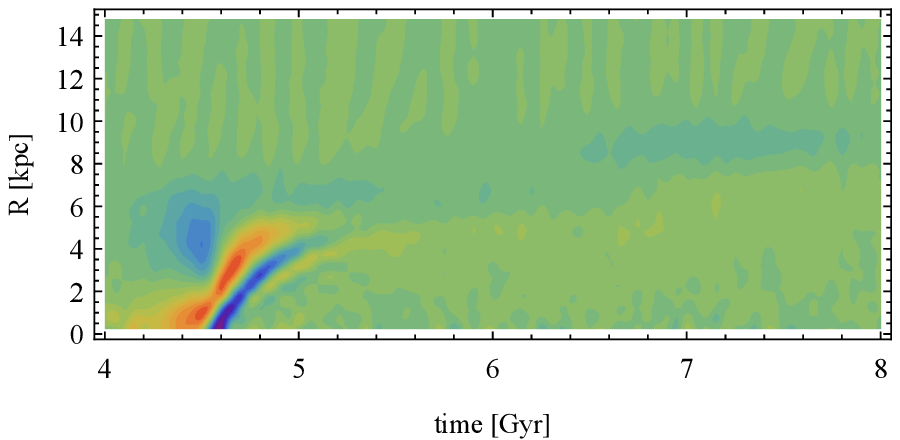}
\\
\vspace{0.3cm}
\hspace{0.7cm} \includegraphics[width=6cm]{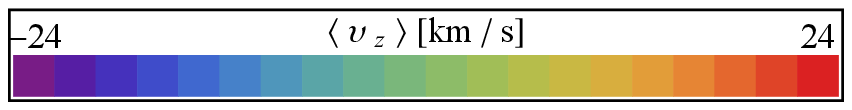}
\includegraphics[width=8.9cm]{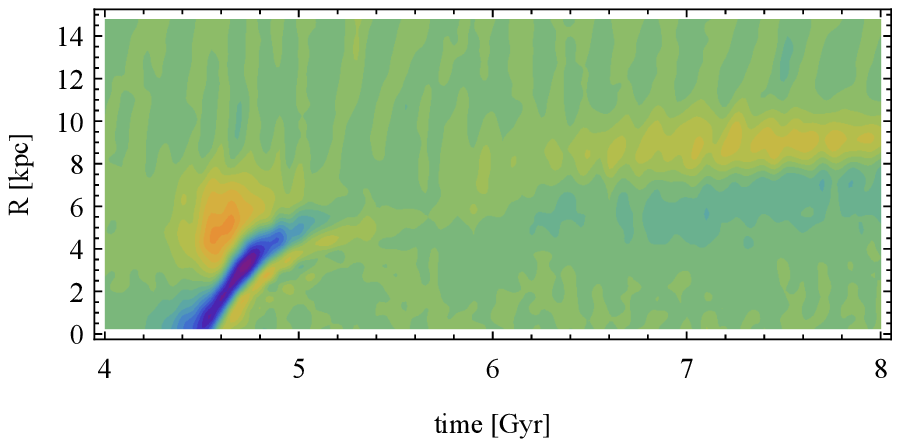}
\\
\vspace{0.3cm}
\hspace{0.7cm} \includegraphics[width=6cm]{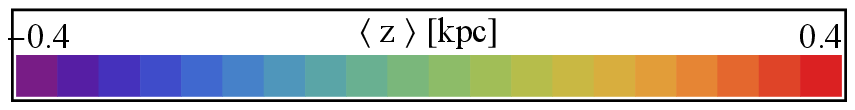}
\includegraphics[width=8.9cm]{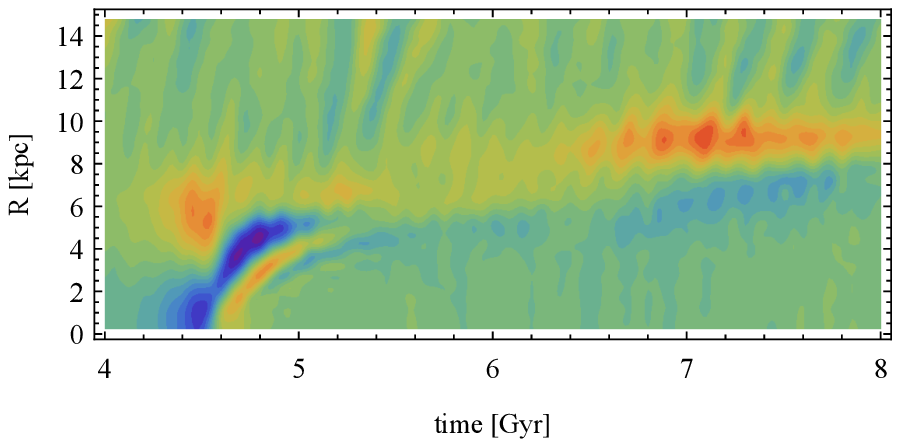}
\caption{Evolution of the profiles of the mean acceleration (upper panel), velocity (middle panel), and distortion of
the positions of the stars (lower panel) along the vertical axis $z$. Positive quantities point along the
angular momentum vector of the disk.}
\label{azvzmeanzprofiles}
\end{figure}

The evolution of the global shape of the  stellar component of the galaxy is illustrated in Fig.~\ref{shape}.
For the purpose
of these and the following measurements the stars within $2 R_{\rm D}$ were used to determine the orientation and
lengths of the principal axes of the stellar distribution in each output and the stellar component was rotated to align
the longest axis with $x$, the intermediate one with $y,$ and the shortest one with $z$ coordinate of a Cartesian
coordinate system that is used throughout the paper. The upper panel of Fig.~\ref{shape} shows the evolution of
the axis ratios $b/a$ and $c/a$ where $a$ is the longest, $b$ the intermediate, and $c$ the shortest axis. In the lower
panel we characterize the shape with a combination of axis ratios in the form of the triaxiality
parameter $T = [1-(b/a)^2]/[1-(c/a)^2]$.

The lower panel of Fig.~\ref{shape} also shows the strength of the bar measured as the $m=2$ mode of the Fourier
decomposition of the surface distribution of stars within $2 R_{\rm D}$ projected along the short axis: $A_m (R) = |
\Sigma_j \exp(i m \theta_j) |/N_s$. Here $\theta_j$ is the azimuthal angle of the $j$th star and the sum is up to the
total number of $N_s$ stars. The radius $R$ is the standard radius in cylindrical coordinates in the plane of the disk,
$R = (x^2 + y^2)^{1/2}$.

The axis ratio $b/a$ decreasing from unity and the growth of the triaxiality parameter $T$ and the bar mode $A_2$ from
zero, characteristic of disks, to much larger values signify the formation of a bar in the stellar component of the
galaxy. The bar grows until $t = 4.3$ Gyr, when a sudden drop occurs in both $T$ and $A_2$ while both $b/a$ and $c/a$
sharply increase. This time marks the occurrence of the buckling event which weakens and thickens the bar.

The bar may be characterized in more detail by calculating the profiles of the bar mode as a function of the
cylindrical radius, $A_2(R)$. A few examples of such profiles at different times are shown in Fig.~\ref{a2profiles}.
They all display a characteristic shape, with a strong growth from zero at small radii, followed by a maximum, and then a
decrease. The radius where $A_2(R)$ drops down to half the maximum value is usually adopted as an estimate of the bar
length. Before the buckling event at $t=4.3$ Gyr, the bar is already quite strong with the maximum of $A_2(R)=0.56$
and the length of the order of 7 kpc. Immediately after buckling, at $t=5$ Gyr, the profile is significantly lower with a
maximum at $A_2(R)=0.45$ and the length also decreased by about 1 kpc. Later on, the bar grows both in strength
and length until $t=10$ Gyr.

\begin{figure*}
\centering
\includegraphics[width=5cm]{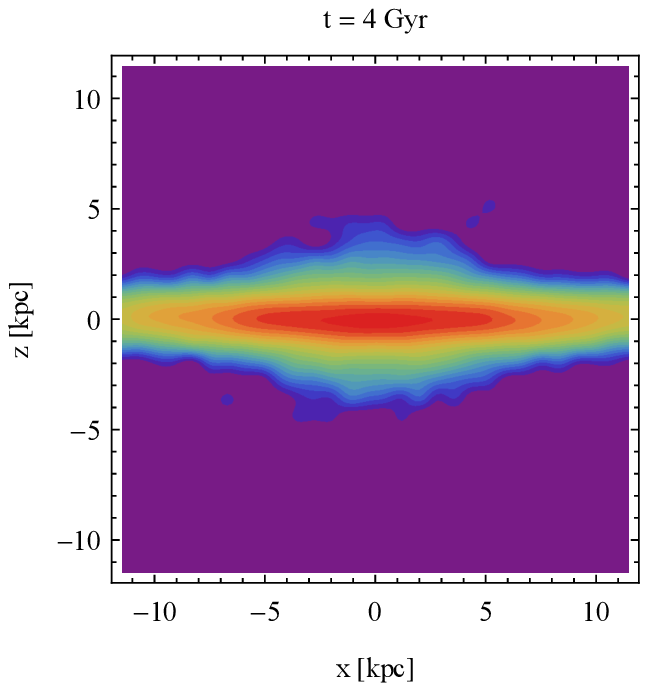}
\includegraphics[width=5cm]{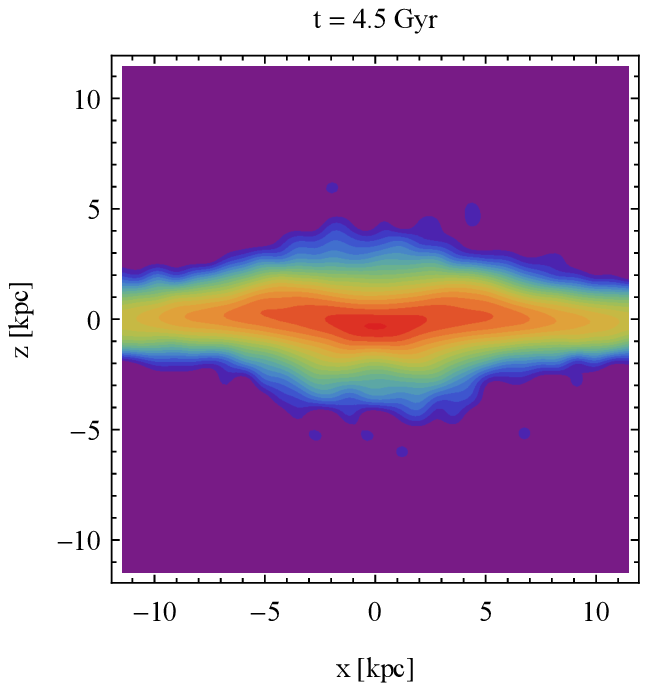}
\includegraphics[width=5cm]{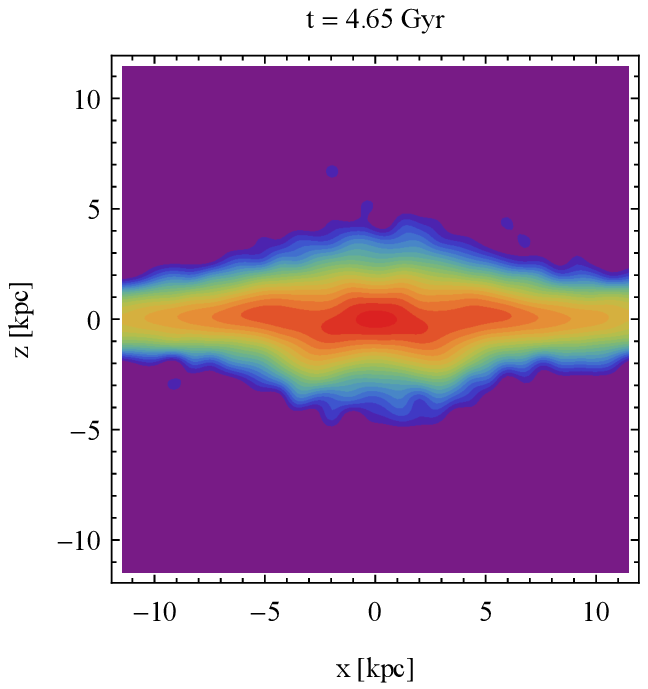} \\
\vspace{0.2cm}
\includegraphics[width=5cm]{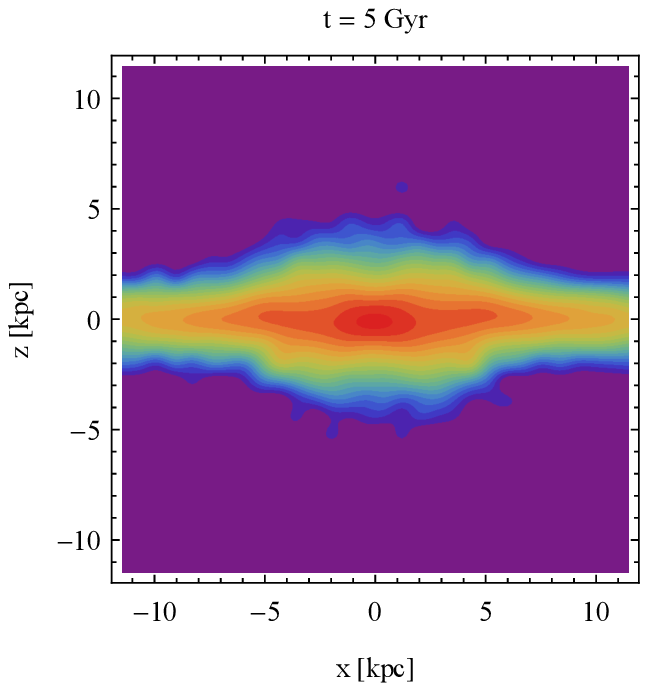}
\includegraphics[width=5cm]{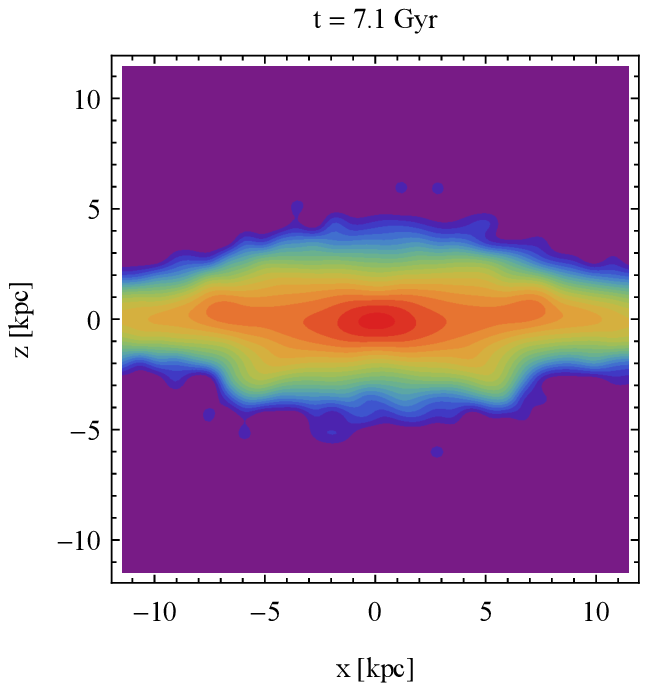}
\includegraphics[width=5cm]{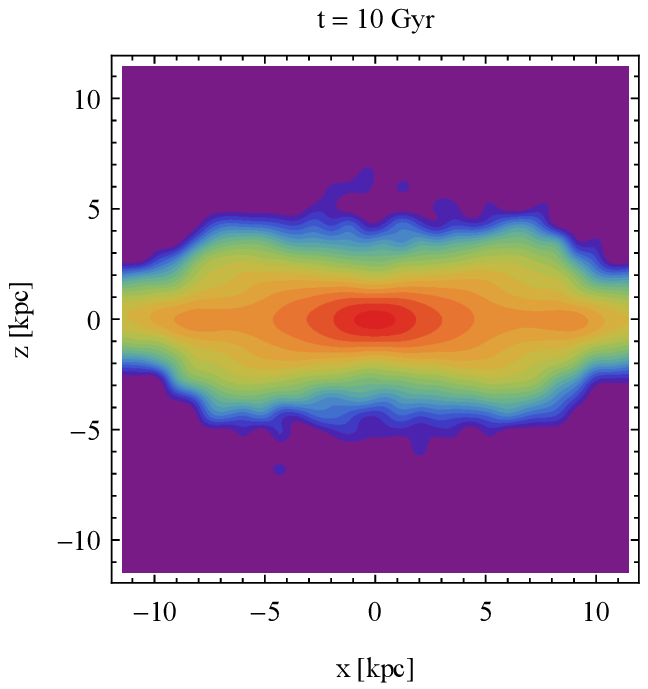} \\
\vspace{0.2cm}
\hspace{0.71cm}
\includegraphics[width=4.2cm]{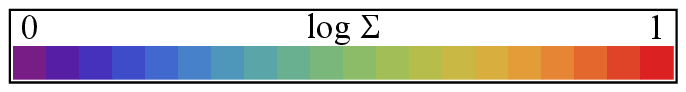}\\
\vspace{0.2cm}
\caption{Surface density distributions of the stellar component viewed edge-on before the first buckling ($t=4$ Gyr), at
the time of first buckling ($t=4.5$, 4.65 and 5 Gyr), at the time of second buckling ($t=7.1$ Gyr), and at the
end of evolution ($t=10$ Gyr). The surface density was normalized to the central maximum value in each case and the
contours are equally spaced in $\log \Sigma$ with $\Delta \log \Sigma = 0.05$.}
\label{surden}
\end{figure*}

The profiles of the bar mode $A_2(R)$ for all outputs can be combined and color-coded to describe the whole history of
the bar evolution in time, as shown in Fig.~\ref{a2modestime}. From this plot it can be seen that the bar starts to form
around $t=1.5$ Gyr in the sense that it crosses the threshold of 0.1 in $A_2(R)$ for the first time. The buckling
around $t=4.5$ Gyr is also visible as a drop of the maximum of $A_2(R)$ from above to below 0.5. At later times the
growth of the bar seems undisturbed, especially after $t=8$ Gyr.

\section{Description of the buckling instability}

Buckling also manifests itself in the evolution of the velocity dispersions of the stars along the bar and
perpendicular to it, which gave rise to its association with the fire-hose instability. In the upper panel of
Fig.~\ref{kinematics} we plot the evolution of the velocity dispersion along the cylindrical radius $\sigma_R$ and
along the vertical direction $\sigma_z$ while the lower panel shows the ratio $\sigma_z/\sigma_R$ as a function of
time. Here again the measurements were done using only the stars within $2 R_{\rm D}$. We can see that the evolution of
$\sigma_R$ follows that of the bar mode $A_2$ in the lower panel of Fig.~\ref{shape} and is due to the fact that as the
bar becomes stronger, more stars are on radial orbits or the orbits are more radial. On the other hand, $\sigma_z$
grows only weakly, similarly to the thickness of the bar. The situation changes abruptly at $t=4.3$ Gyr when $\sigma_R$
suddenly drops down while $\sigma_z$ grows. The event is even more emphasized in the ratio $\sigma_z/\sigma_R$ shown in
the lower panel of Fig.~\ref{kinematics}. It has been suggested in the past that the low value of $\sigma_z/\sigma_R$
immediately before buckling is the reason for its occurrence because it leads to a kind of fire-hose instability which
moves the stars out of the disk plane and then leaves the bar thickened. We note, however, that the minimum value of
$\sigma_z/\sigma_R$ immediately before buckling is much higher than the canonical value of 0.3 derived theoretically
for some idealized configurations and thus speaks against this interpretation. Therefore, it is possible that the
departures of the stars out of the disk plane are due to resonances between the radial and vertical motions.

\begin{figure*}
\centering
\includegraphics[width=5cm]{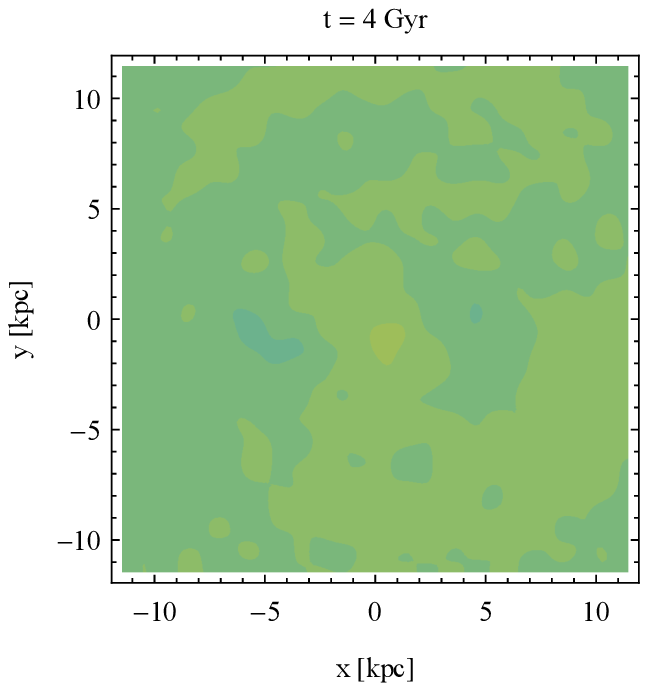}
\includegraphics[width=5cm]{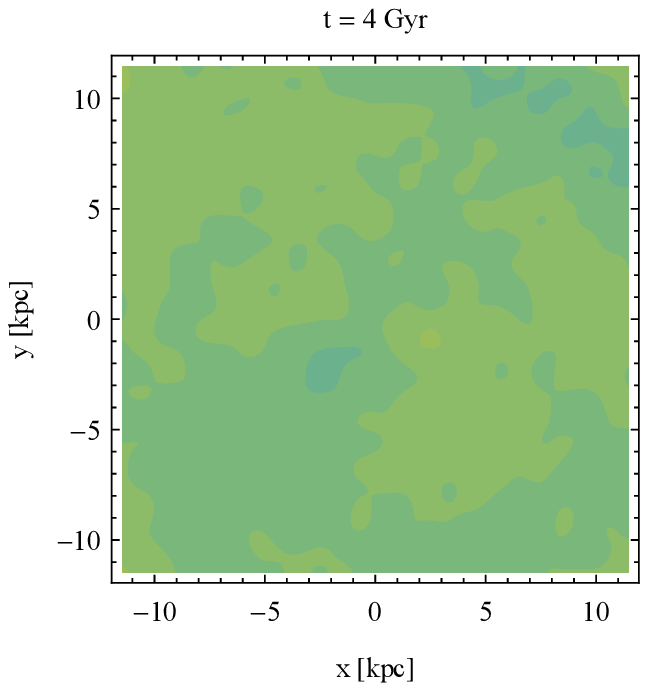}
\includegraphics[width=5cm]{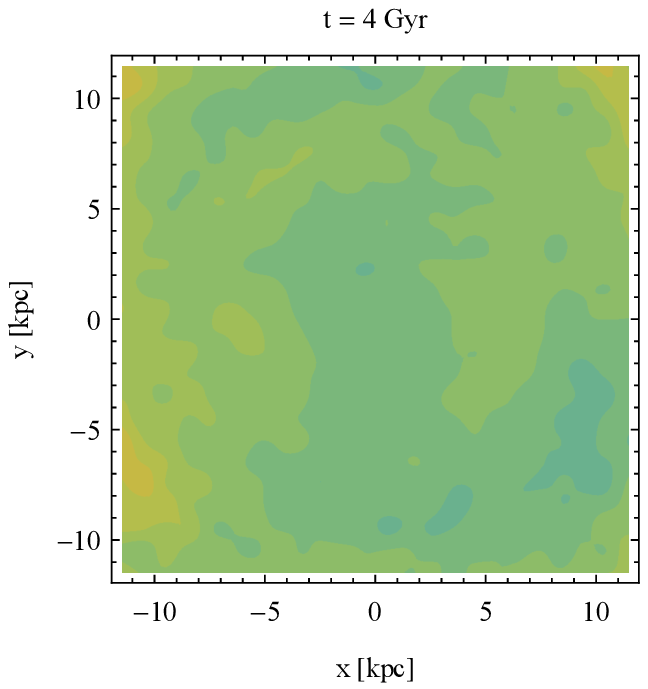}\\
\vspace{0.2cm}
\includegraphics[width=5cm]{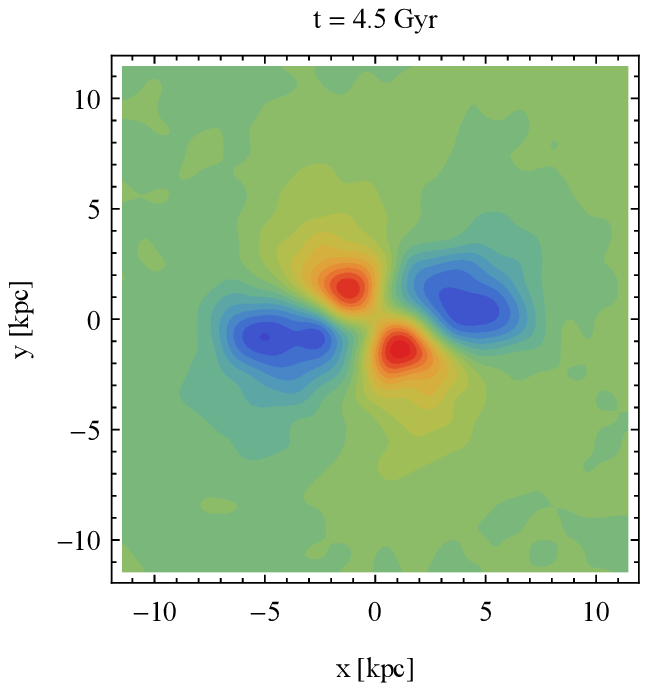}
\includegraphics[width=5cm]{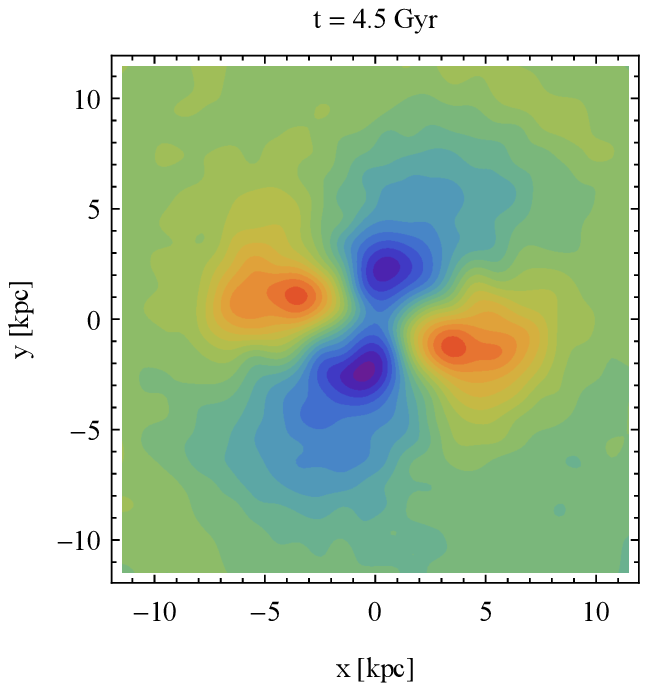}
\includegraphics[width=5cm]{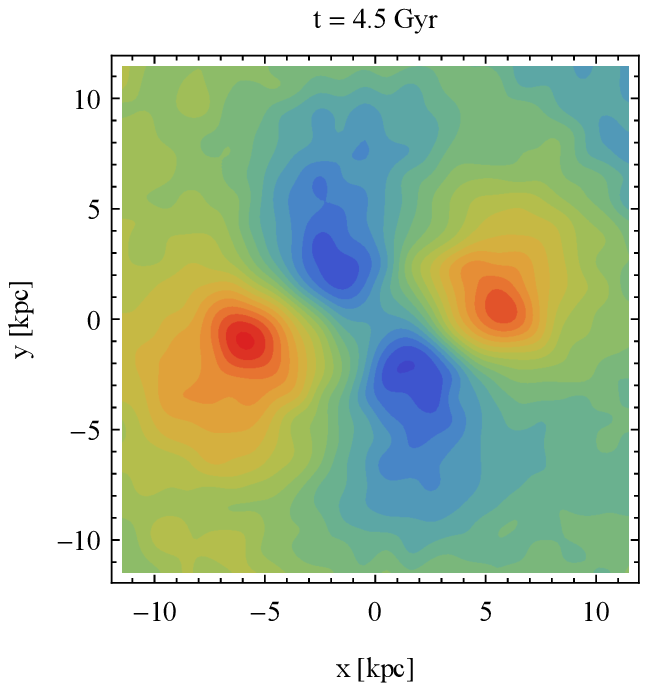}\\
\vspace{0.2cm}
\includegraphics[width=5cm]{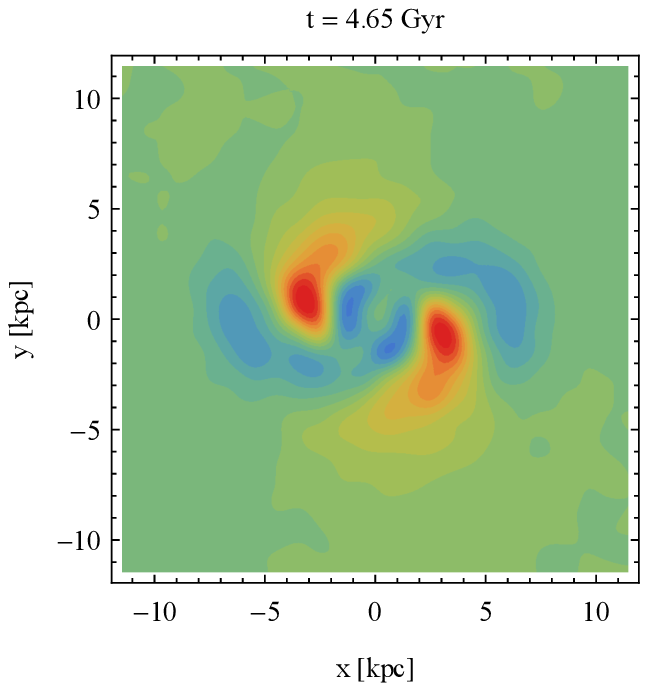}
\includegraphics[width=5cm]{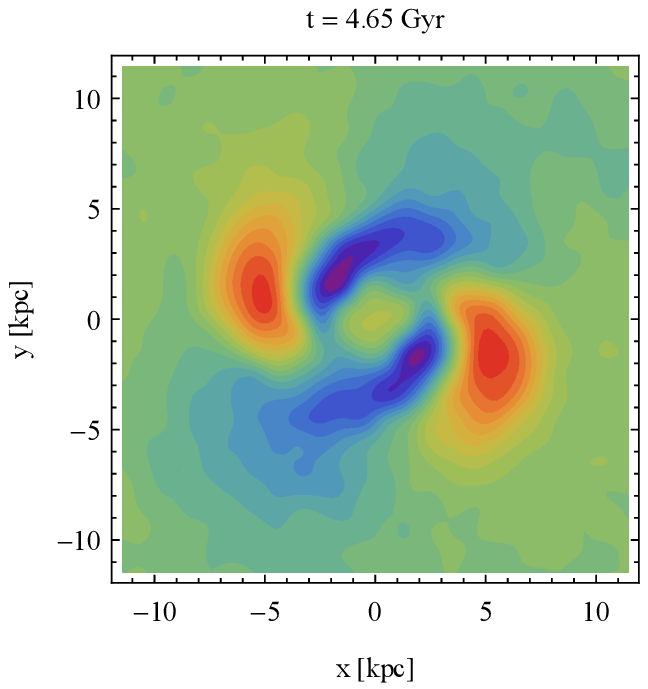}
\includegraphics[width=5cm]{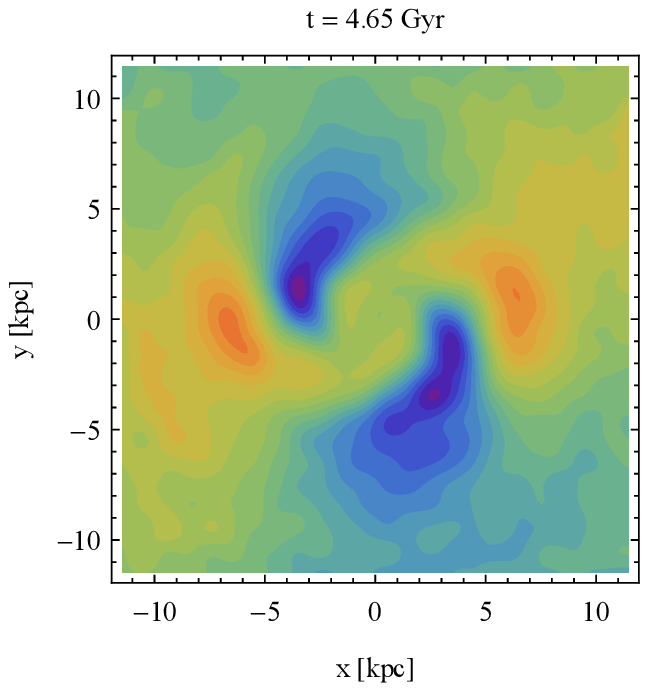}\\
\vspace{0.2cm}
\includegraphics[width=5cm]{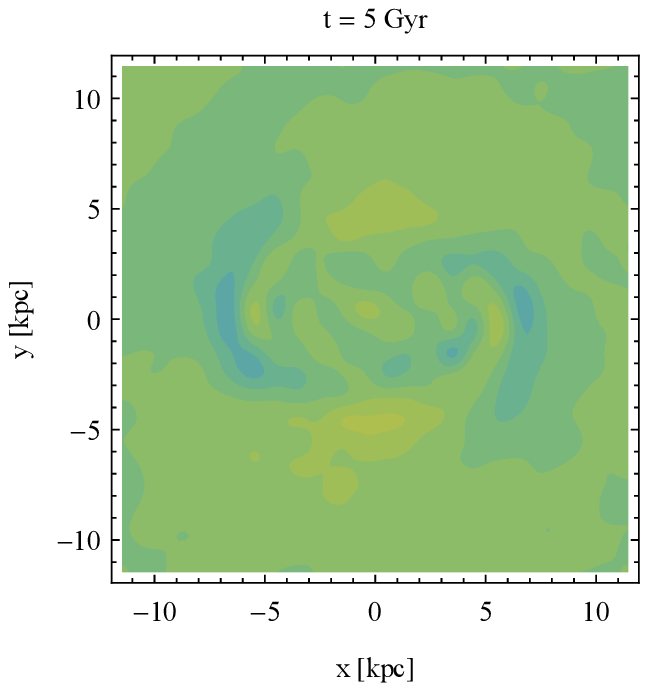}
\includegraphics[width=5cm]{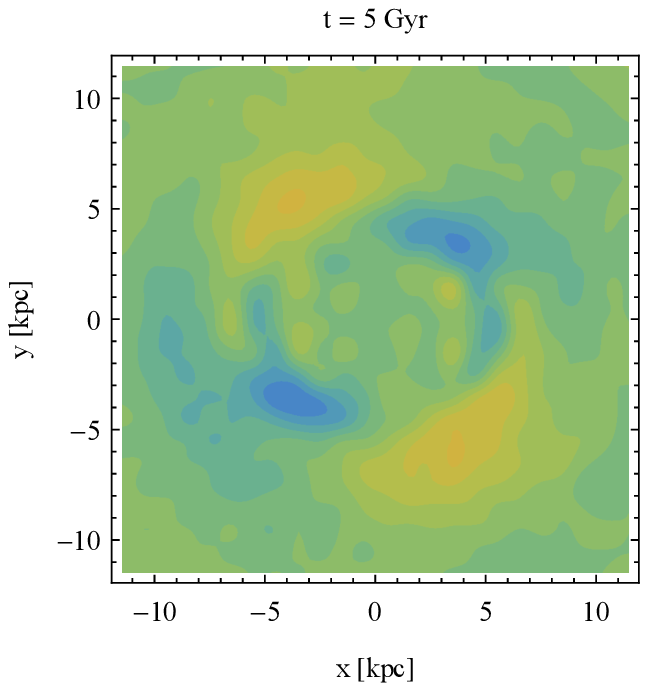}
\includegraphics[width=5cm]{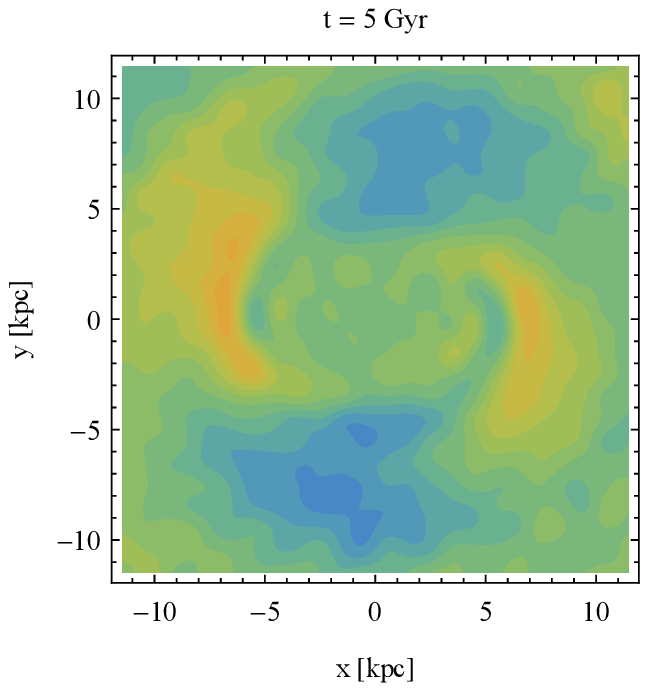}\\
\vspace{0.2cm}
\hspace{0.7cm}
\includegraphics[width=4.2cm]{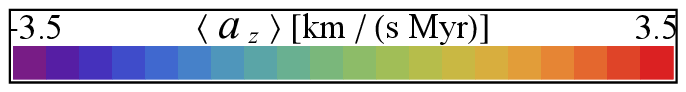}
\hspace{0.71cm}
\includegraphics[width=4.2cm]{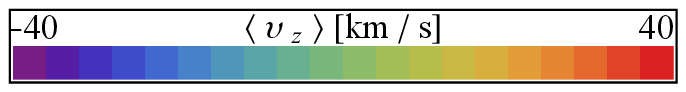}
\hspace{0.71cm}
\includegraphics[width=4.2cm]{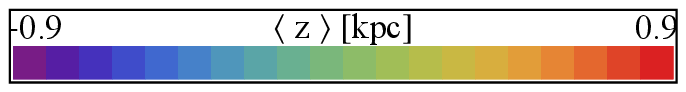}\\
\vspace{0.2cm}
\caption{
Face-on maps of the mean acceleration (left column), mean velocity (middle column),
and mean distortion of the positions of the stars (right column) along the vertical direction at the time of the first
buckling. The four rows of panels show the results for times $t=4$, 4.5, 4.65, and 5 Gyr. }
\label{azvzmeanz1}
\end{figure*}

\begin{figure*}
\centering
\includegraphics[width=5cm]{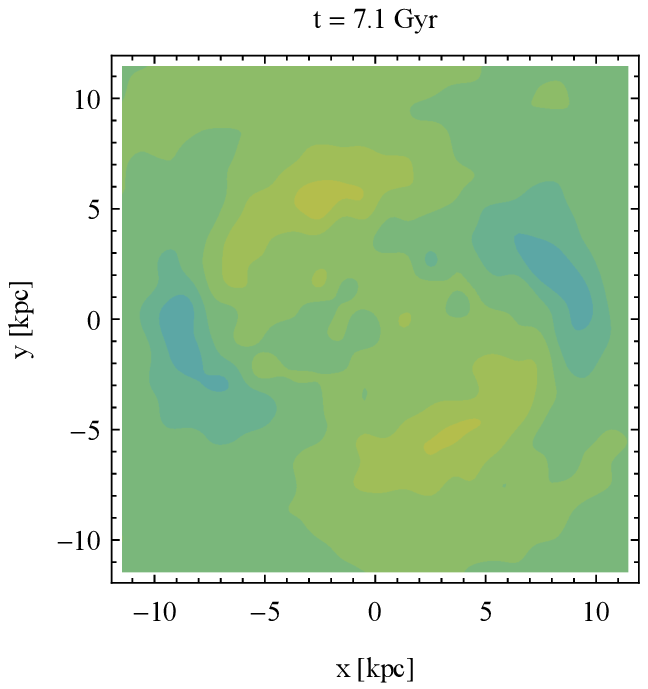}
\includegraphics[width=5cm]{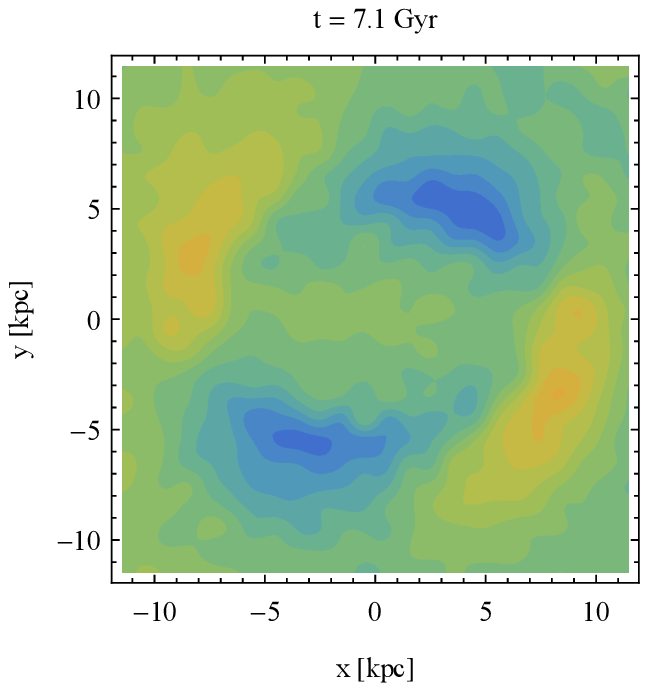}
\includegraphics[width=5cm]{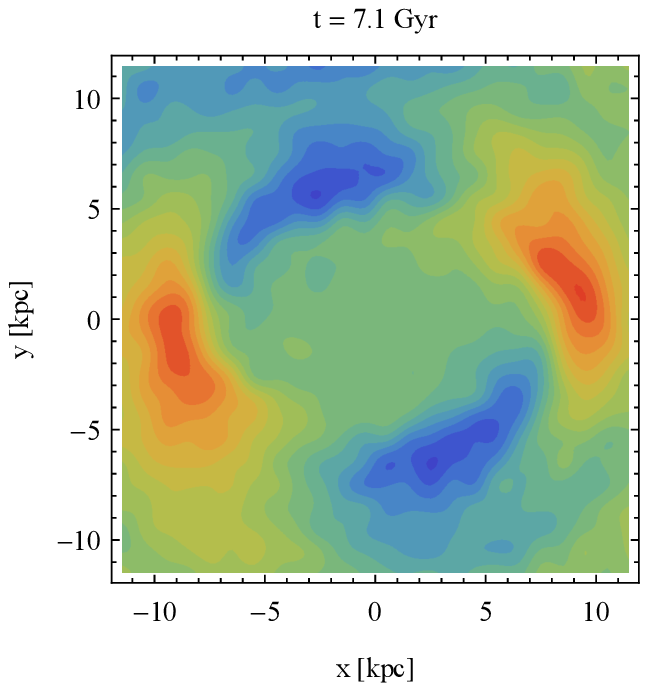}\\
\vspace{0.2cm}
\includegraphics[width=5cm]{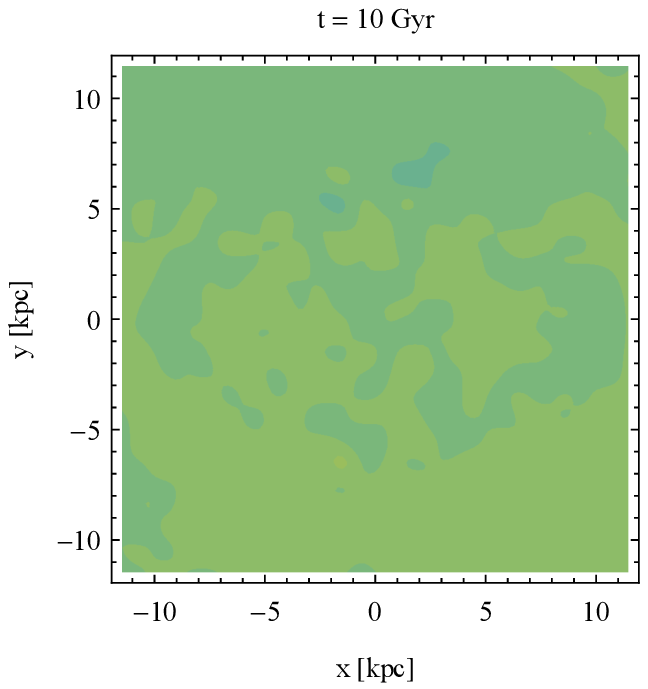}
\includegraphics[width=5cm]{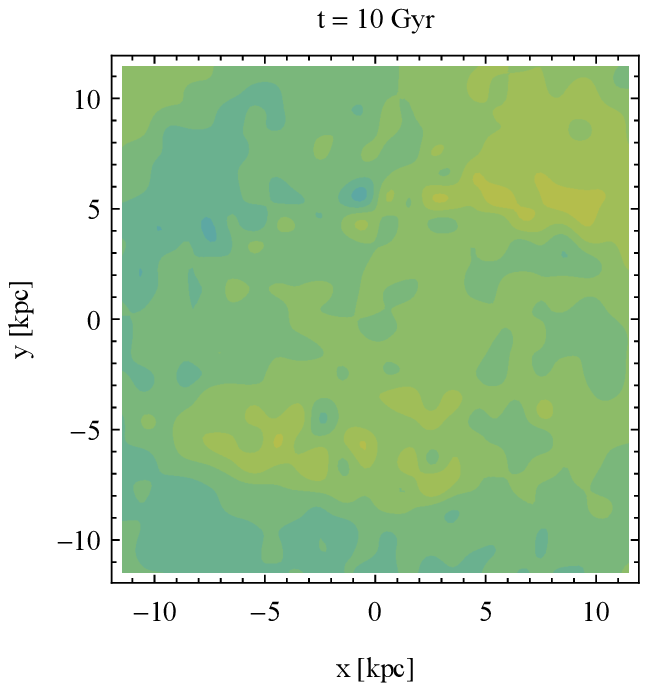}
\includegraphics[width=5cm]{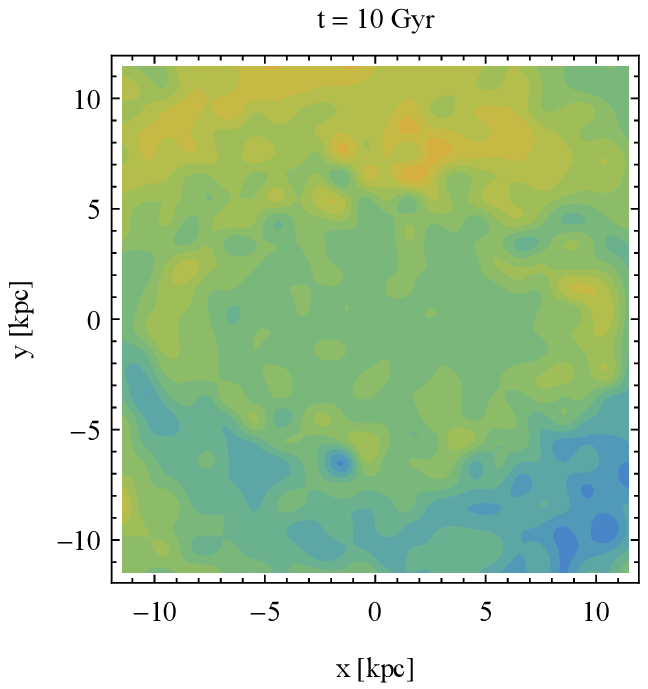}\\
\vspace{0.2cm}
\hspace{0.7cm}
\includegraphics[width=4.2cm]{legend18.eps}
\hspace{0.71cm}
\includegraphics[width=4.2cm]{legend15.eps}
\hspace{0.71cm}
\includegraphics[width=4.2cm]{legend14.eps}\\
\vspace{0.2cm}
\caption{Same as Fig.~\ref{azvzmeanz1} but at the time of second buckling (upper row) and at the end of evolution
(lower row).}
\label{azvzmeanz2}
\end{figure*}

While the dispersions only measure the random motion of the stars in a given direction, it is much more instructive to
study the acceleration and the streaming velocity of the stars as well as their distortion along the vertical
direction. One way to do this is to calculate these quantities in shells along the cylindrical
radius. The results of such measurements in terms of the profiles of the mean acceleration $\langle a_z \rangle$, mean
velocity $\langle \upsilon_z \rangle,$ and the mean distortion of the positions of the stars along the vertical axis
$\langle z \rangle$ as a function of time are shown color-coded in Fig.~\ref{azvzmeanzprofiles} in the three
panels, respectively. The time range covered in these plots was restricted to between $t=4$ and $t=8$
Gyr when the signal is nonzero. We immediately see a strong buckling event with a maximum at $t=4.5$ Gyr starting with
a smile-like distortion of the bar which very soon after, around $t=4.6$ Gyr, changes into a frown-like one (with
respect to the direction of the  angular momentum  vector of the galaxy). After this the distortion propagates outward and
becomes very weak already at $t=5$ Gyr. We note that, as expected, the direction of the acceleration is
opposite to the distortion and the signal in the acceleration starts to be visible at earlier times.

Interestingly, later on, between $t=6.5$ and $t=8$ Gyr, a second buckling event takes place. This happens only at
larger radii, $R > 5$ kpc, lasts much longer, and involves only one kind of distortion (downward then upward with
radius), without a clear reversal that was present in the earlier episode of buckling. While no obvious signal of this
buckling is seen in the evolution of the bar mode in Fig.~\ref{a2modestime}, a hint of its presence can be seen in the
evolution of the axis ratio $c/a$ in the upper panel of Fig.~\ref{shape}, which increases around this time;
albeit very slightly because the measurements there were done within $2 R_{\rm D} = 6$ kpc. We note that there is even
a kind of continuity between the first and second buckling events as the weak distortion is present all the time
between $t=5$ and $t=6.5$ Gyr. In this sense the first buckling event may be considered as a seed for the second one.
Some examples of the edge-on view of the galaxy at different times during buckling are shown in Fig.~\ref{surden}.

Even more insight into the structure of the bar during buckling can be obtained by plotting the maps of the mean
vertical acceleration, velocity, and distortion in the face-on view. A few examples of such maps are shown in
Figs.~\ref{azvzmeanz1} and \ref{azvzmeanz2} for the same outputs as the corresponding surface density
plots in the edge-on view in Fig.~\ref{surden}. In all plots the stellar component has been rotated so that the bar is
aligned with the $x$ axis, while its shortest axis (and the rotation axis of the disk) is along $z$ and the disk is
rotating counterclockwise.

In the first row of panels in Fig.~\ref{azvzmeanz1}, corresponding to $t=4$ Gyr, no clear signal in the acceleration,
velocity, or distortion is visible yet, in agreement with Fig.~\ref{azvzmeanzprofiles}. In the second row, at $t=4.5$,
a very strong signal is seen. We note that the regions of downward acceleration (in blue) and upward velocity and
distortion (in red) are not located exactly along the bar (which is aligned with the $x$ coordinate) and the regions of
upward acceleration (in red) and downward velocity and distortion (in blue) are not exactly perpendicular to it. At the
same time, the patterns in distortion and velocity are rotated with respect to each other by about $45$ deg so that the
regions of extreme distortion are aligned with the zero-velocity curves and vice versa, while the maxima of
acceleration coincide with the maxima of distortion but have an opposite sign, as expected for an oscillatory motion
in the vertical direction.

This growing pattern in acceleration, velocity, and distortion preserves its orientation with respect to the
bar for about 0.2 Gyr, that is between $t=4.4$ and $t=4.6$. After this short period it starts to wind up, as shown in
the third row of plots in Fig.~\ref{azvzmeanz1}. Later, at $t=5$ Gyr (lower row of panels in
Fig.~\ref{azvzmeanz1}), a weak pattern remains only in the outer parts of the bar ($R > 5$ kpc) while within the inner
range of radii ($R < 5$ kpc) a clear boxy/peanut shape is formed (see Fig.~\ref{surden}).

Figure~\ref{azvzmeanz2} illustrates the second buckling event that occurs in the outer parts of the bar ($R > 5$
kpc) between $t=6.5$ and $t=8$ Gyr. The upper set of panels shows the acceleration, velocity, and distortion maps at $t =
7.1$ Gyr when the buckling signal is strongest. A similar pattern in $\langle a_z \rangle$, $\langle \upsilon_z
\rangle,$ and $\langle z \rangle$ is discernible, but only in the outskirts of the bar, while its inner part remains
undisturbed. The origin of these distortions can be seen in the corresponding panel of Fig.~\ref{surden} showing the
surface density distribution of the stars. Clearly, there is a frown-like distortion dominating around $R=6$ kpc and a
smile-like one around $R=8$ kpc. Later on, the pattern in all quantities dissipates leaving behind at the end
of evolution ($t=10$ Gyr) a much bigger boxy/peanut shape (see the lower right panel of Fig.~\ref{surden}).

\begin{figure*}
\centering
\includegraphics[width=15cm]{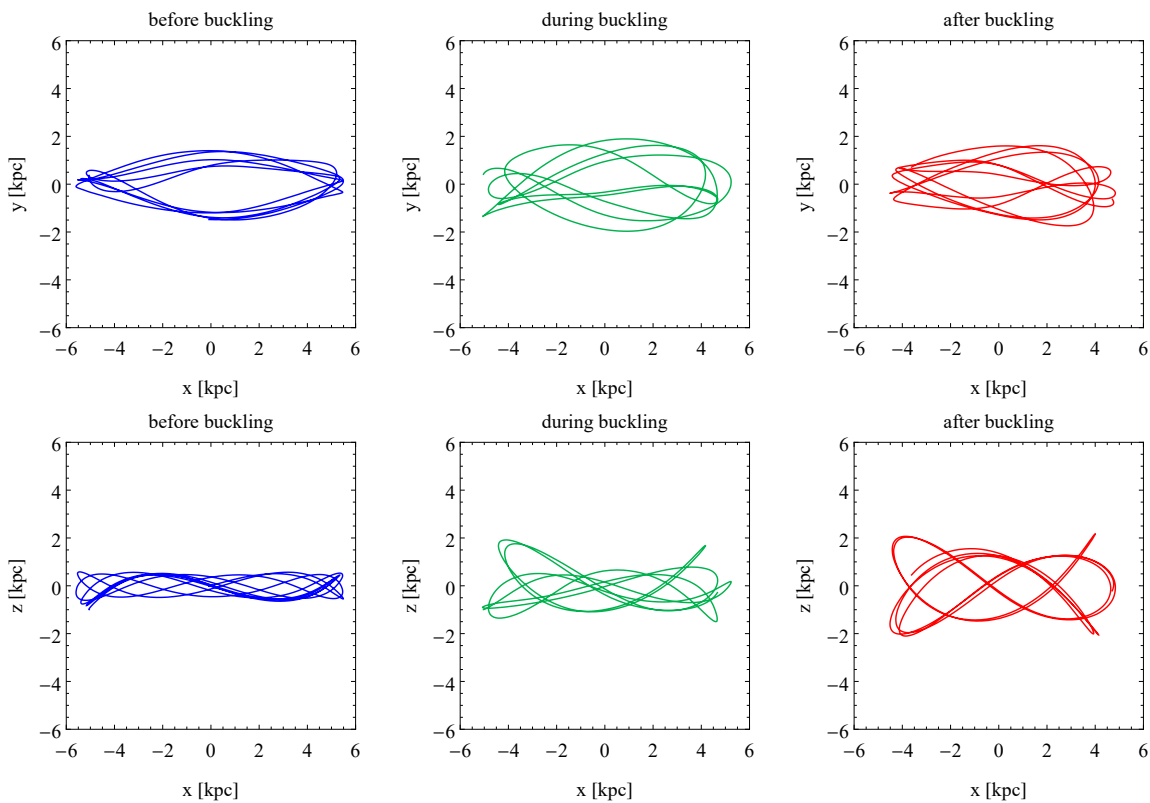}
\caption{Example of a buckling orbit. The orbit is plotted in different colors in three time periods: before, during,
and after buckling (from the left to the right panel) and in two projections: in the face-on and edge-on view (the top
and bottom row, respectively). After buckling the star follows a regular, periodic, pretzel-like orbit with
frequency ratio $f_{z, {\rm \: after}}/f_{x, {\rm \: after}} =7/4$, that is it completes four oscillations in $x$ for
every seven oscillations in $z$.}
\label{orbit80}
\end{figure*}

\begin{figure}
\centering
\includegraphics[width=8.9cm]{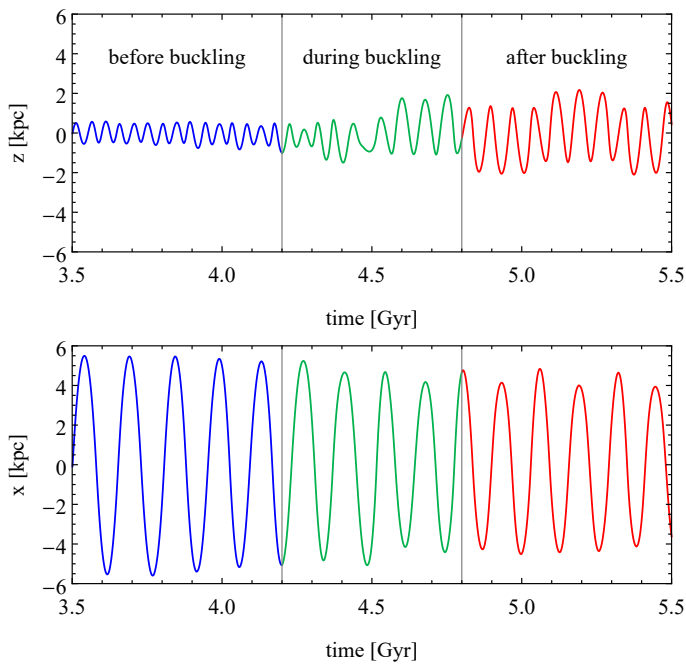}
\caption{Evolution of the position of the star along the minor and major axes of the bar for the orbit shown in
Fig.~\ref{orbit80}. The time range of interest was divided into three parts: before, during, and after buckling,
indicated with different colors.}
\label{orbit80time}
\end{figure}

\section{The orbital structure of the buckling bar}

To obtain a deeper understanding of the buckling instability it seems indispensable to resort to the study of the orbital
structure of the buckling bar. In this section we restrict the analysis of the orbits to the first, stronger buckling
episode. For this purpose we reran the simulation in the 2 Gyr period between $t=3.5$ and $t=5.5$ Gyr saving 2001
outputs, that is, every 0.001 Gyr. This time resolution is sufficient to measure the properties of orbits for a majority
of stars building up the bar. We also determined the orientation of the bar in each simulation output and transformed
the orbits to the reference frame of the bar. We thus study the orbits in the Cartesian reference frame with $x$, $y,$
and $z$ aligned with the major, intermediate, and minor axes of the bar. The advantages of this approach in comparison
to using the traditional cylindrical coordinates have been discussed in detail by \citet{Valluri2016} and
\citet{Gajda2016}.

\begin{figure*}
\centering
\includegraphics[width=15cm]{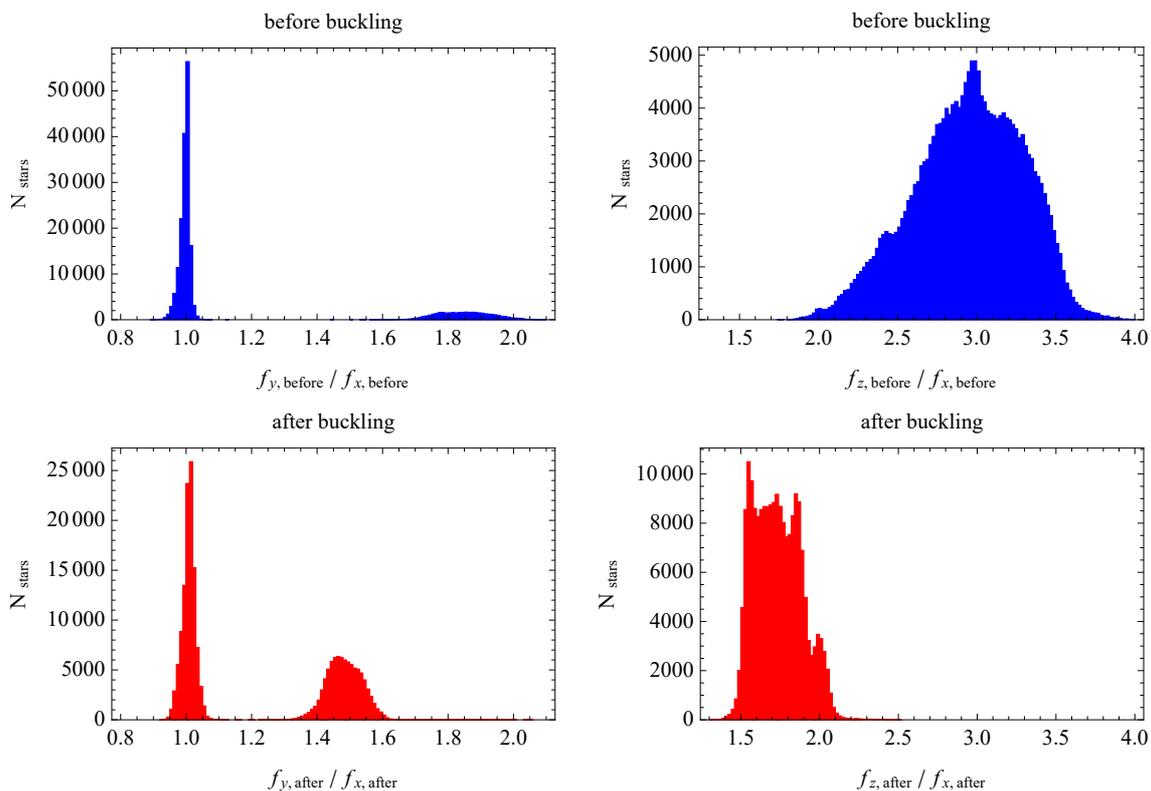}
\caption{Histograms of frequency ratios $f_y/f_x$ (left column) and $f_z/f_x$ (right column) before (upper row, blue)
and after (lower row, red) buckling.}
\label{histograms}
\end{figure*}

A visual inspection of a few hundred orbits revealed that an abrupt change in
their properties takes place around $t=4.5$ Gyr, as suggested by the results of the previous section. We therefore
divided this 2 Gyr period into three parts which we refer to as `before buckling', `during buckling', and `after
buckling'. The first period of 0.7 Gyr falls between 3.5 and 4.2 Gyr, the second one lasting 0.6 Gyr is between 4.2 and
4.8 Gyr (and is therefore centered on the 4.5 Gyr time when the transition takes place) and the third one again lasting
0.7 Gyr is between 4.8 and 5.5 Gyr. It is certainly possible to measure the properties of the orbits when they are
stable and therefore we take measurements only in the two periods of 0.7 Gyr before and after buckling.

Figure~\ref{orbit80} shows an example of a stellar orbit that experiences buckling. Clearly the shape of the orbit is
very different at the beginning; in particular its motion in the $z$ direction (perpendicular to the bar) is confined
to a very narrow range (left panels). During buckling the shape of the orbit experiences an abrupt change (middle panels)
so that after buckling (right panels) the orbit covers a much larger range of $z$ values. The oscillatory motion of the
star in the directions perpendicular and parallel to the bar is shown in Fig.~\ref{orbit80time}. Clearly, both the
amplitude and the frequency of the motion
change in each of the $x$ and $z$ directions.

We use time series such as the ones shown in Fig.~\ref{orbit80time} to calculate the frequencies of the orbits in
the Cartesian coordinates $x$, $y,$ and $z$ using discrete Fourier transform. The calculations were done for the 0.7 Gyr
time-periods before and after buckling that include 700 data points each. This time range needs to be
long enough for the orbits to complete at least a few oscillations. On the other hand, we recall that the measurements
of orbital properties here are done `in vivo', that is in the live evolving bar, and therefore the time period considered should
be as short as possible so that the bar properties can be approximated as constant. In order to obtain sufficiently accurate
estimates of frequency we restrict the analysis to orbits with periods larger than 0.25 Gyr. For each orbit we
also estimate the amplitude (or apocenter) in the three coordinates and require that the position of the star crosses the
zero value (in the reference frame of the bar) at least once. These conditions eliminate the large nearly circular
orbits that do not contribute to the bar. Such weak restrictions allow us to measure the properties of 475512 out of
the total number of $10^6$ stars.

Since we are interested in the orbits that contribute to the buckling of the bar we further restrict the analysis to
orbits that actually support the bar, that is: their $x$ amplitude, $a_x$, is within 7 kpc (the length of the bar) and
their amplitude in the $y$ direction, $a_y$, is smaller than 0.7 of the $x$ amplitude: $a_y < 0.7 a_x$. We require that
these conditions are fulfilled both before and after buckling. This way we eliminate for example x2 and x4 orbits that
do not support the bar, as well as long-axis tubes in the center. In addition, we require that the orbits actually
buckle, and therefore we take only those orbits that have larger amplitudes in $z$ after buckling than before buckling: $a_{z, {\rm \:
after}} > a_{z, {\rm \: before}}$. These restrictions leave us with a sample of 199601 orbits whose properties we study
further.

A common way to characterize the orbital structure of a bar is to study the distribution of the frequency ratios. In
Fig.~\ref{histograms} we show histograms of frequency ratios $f_y/f_x$ and $f_z/f_x$ before and after buckling. In the
case of $f_y/f_x$ most of the orbits contribute to the strong peak at $f_y/f_x = 1$ characteristic for x1 orbits, both
before and after buckling. There is also an additional small contribution from other, box orbits with initial
frequencies in the range 1.7-2.0. These orbits shift to values close to 3/2 after buckling although some of
the x1 orbits also contribute to this new peak, since this range is more populated than the 1.7-2.0 range before
buckling.

\begin{figure}
\centering
\includegraphics[width=7.7cm]{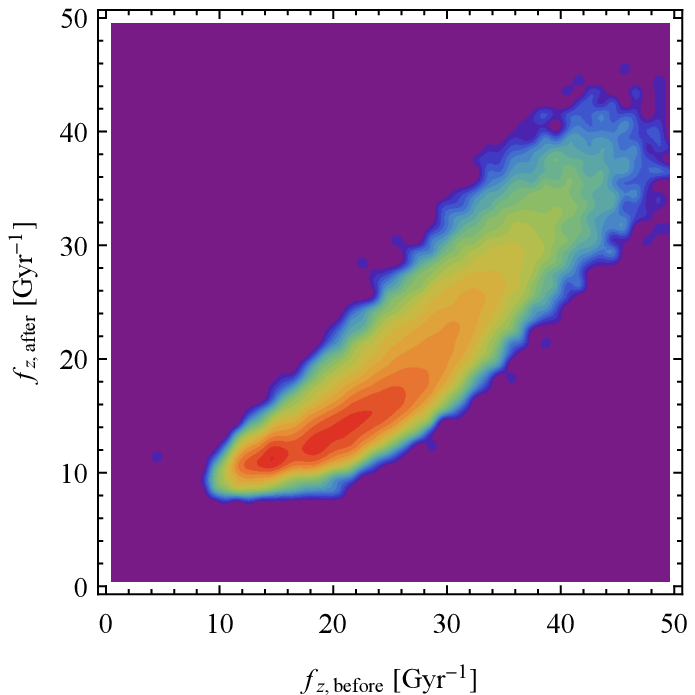} \\
\vspace{0.2cm}
\includegraphics[width=7.7cm]{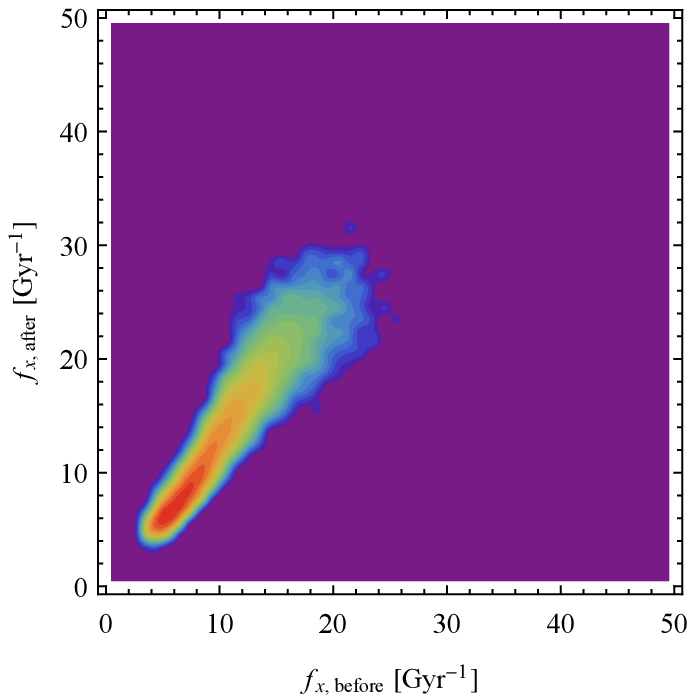}
\caption{Distribution of the stars in the plane of the frequency of the orbit along a given direction after buckling vs.
the same frequency before buckling. The upper panel shows the results for the motion along $z$ and the lower panel for
the motion along $x$. The color coding in these 2D histograms and the following figures is such that the red
corresponds to maximum occupation and violet to empty cells.}
\label{freqzfreqx}
\end{figure}

\begin{figure}
\centering
\includegraphics[width=7.7cm]{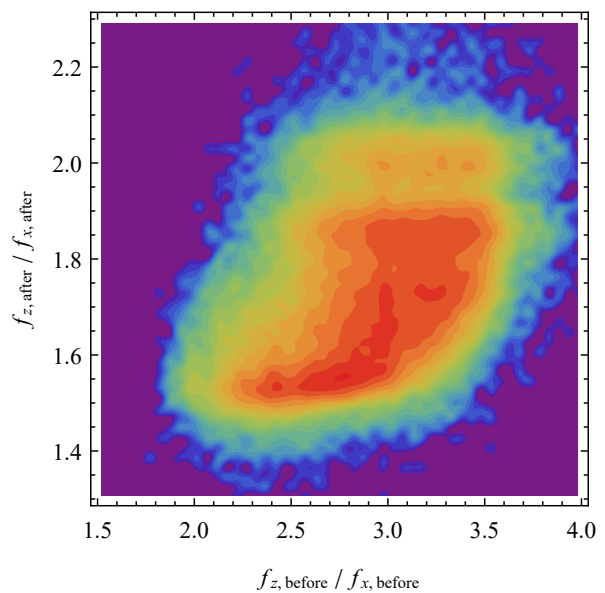}
\caption{Distribution of stars in the plane of frequency ratios $f_{z, {\rm \: before}}/f_{x,
{\rm \: before}}$ and $f_{z, {\rm \: after}}/f_{x, {\rm \: after}}$.}
\label{freqzfreqxbeforeafter}
\end{figure}

A much more interesting evolution occurs in the ratio $f_z/f_x$. Before buckling, the orbits occupy a wide range of
values between 1.8 and 4. After buckling they move to a much narrower range of much smaller values: 1.4-2.1. The reason
for this can be partially understood by looking at the evolution of $f_z$ and $f_x$ separately, which is shown in
Fig.~\ref{freqzfreqx}. We can see that while the frequencies in the vertical direction, $f_z$, typically decrease
during buckling, those along the bar, $f_x$, increase. The effect of this is obviously to strongly decrease the ratio
$f_z/f_x$. Another way to look at this change is via a 2D histogram of the ratio $f_z/f_x$ before and after buckling
shown in Fig.~\ref{freqzfreqxbeforeafter}. This plot can be helpful in mapping between $f_{z, {\rm \: before}}/f_{x,
{\rm \: before}}$ and $f_{z, {\rm \: after}}/f_{x, {\rm \: after}}$. The right-hand histograms of Fig.~\ref{histograms}
can be viewed as projections of the 2D histogram in Fig.~\ref{freqzfreqxbeforeafter} in the vertical and horizontal
directions.

Let us see how the orbits evolve in terms of frequency using the particular example of the orbit shown in
Fig.~\ref{orbit80}. Before buckling, the frequency ratios for this orbit are $f_{y, {\rm \: before}}/f_{x,
{\rm \: before}} = 1$ and $f_{z, {\rm \: before}}/f_{x, {\rm \: before}} = 3$, which place the orbit close to the
highest peaks in the histograms of the upper row in Fig.~\ref{histograms}. After buckling, the frequency ratios change
to $f_{y, {\rm \: after}}/f_{x, {\rm \: after}} = 3/2$ and $f_{z, {\rm \: after}}/f_{x, {\rm \: after}} = 7/4=1.75$.
This means that in terms of $f_y/f_x$ the orbit shifted to the second, smaller peak, while in terms of $f_z/f_x$ it
shifted to the second peak from the left after buckling.

We note that the shape of the $f_z/f_x$ histogram after buckling is similar to those found by \citet{Portail2015} for
a number of Milky Way-like models with a boxy/peanut bar. In particular, our histogram, like one of their models, has a
strong peak around $f_{z, {\rm \: after}}/f_{x, {\rm \: after}} = 1.5$ and a smaller peak near $f_{z, {\rm \:
after}}/f_{x, {\rm \: after}} = 2$ corresponding to banana-like orbits. There are also intermediate peaks around 1.75
and 1.85, the first of which is represented by our example orbit. In the orbit classification of \citet{Portail2015}
who named orbits A-F based on the $f_z/f_x$ ratio from 1.5 to 2.0 (see their fig. 1) our example orbit with
$f_z/f_x=1.75$ falls between classes C and D and is different from all their example orbits (see their fig. 2).

\begin{figure*}
\centering
\hspace{0.11cm}
\includegraphics[width=6.77cm]{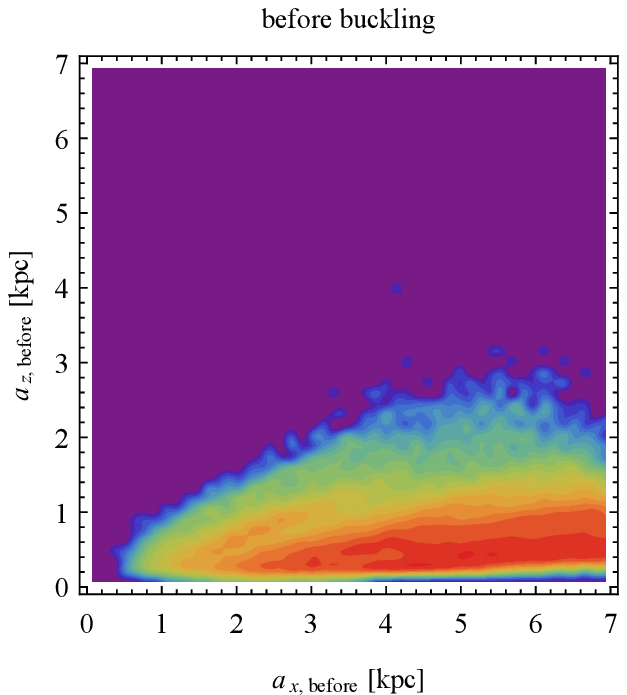}
\hspace{0.15cm}
\includegraphics[width=6.77cm]{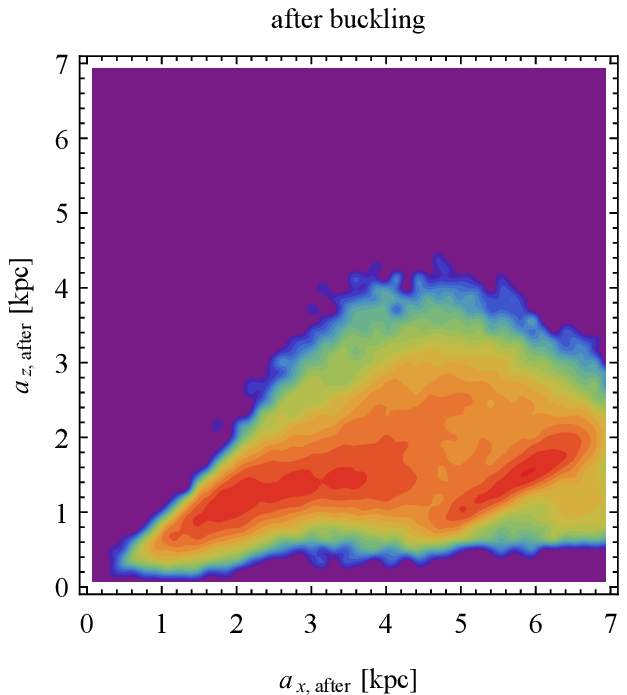} \\
\vspace{0.3cm}
\includegraphics[width=7.cm]{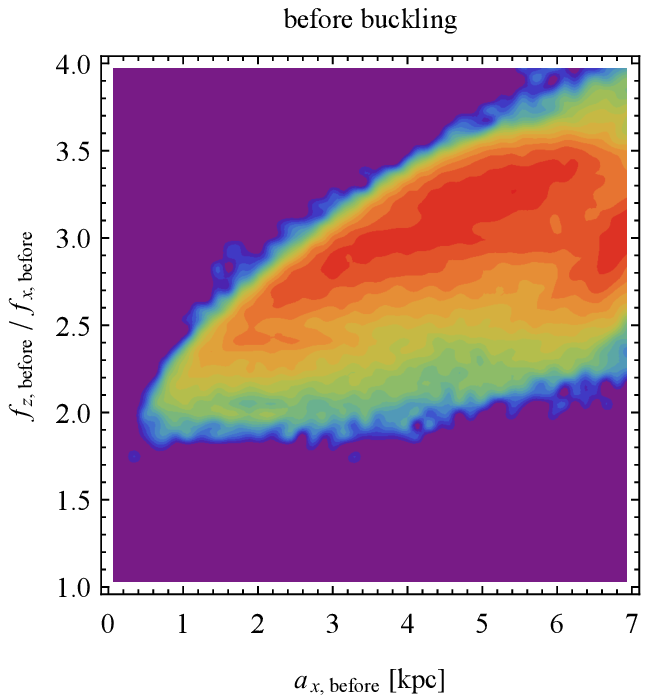}
\includegraphics[width=7.cm]{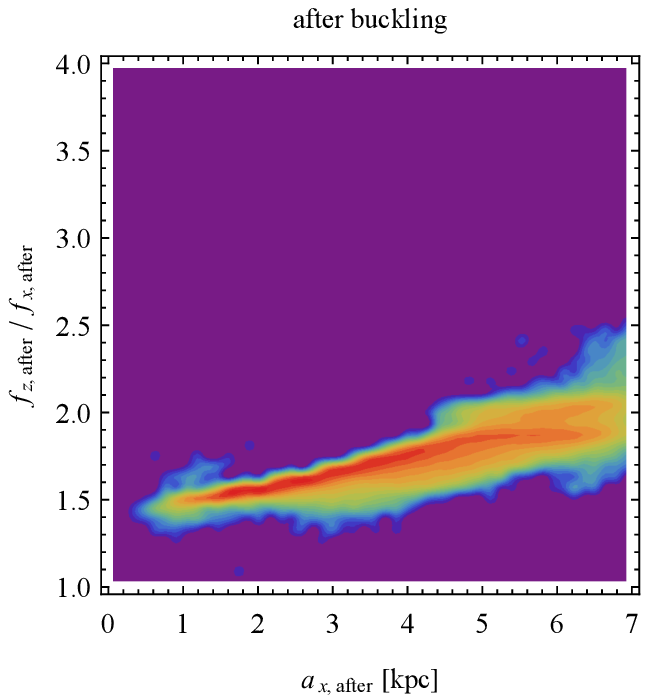} \\
\vspace{0.3cm}
\hspace{0.01cm}
\includegraphics[width=7.cm]{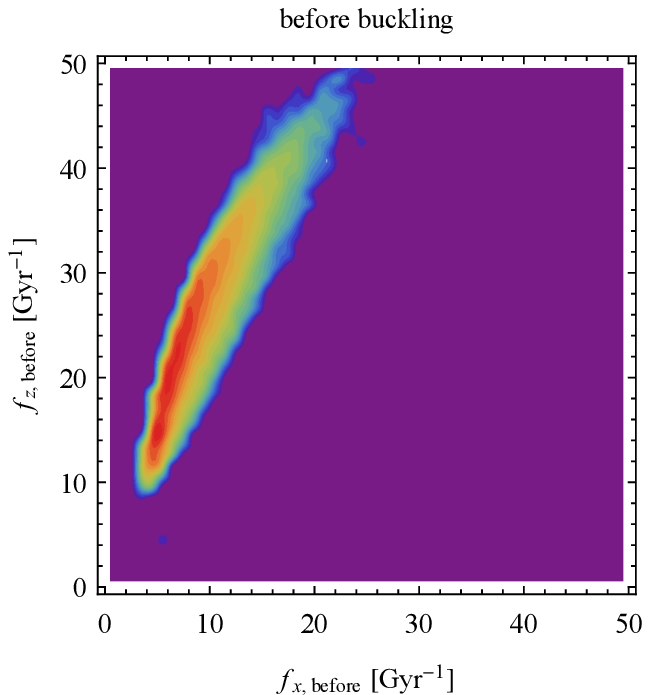}
\includegraphics[width=7.cm]{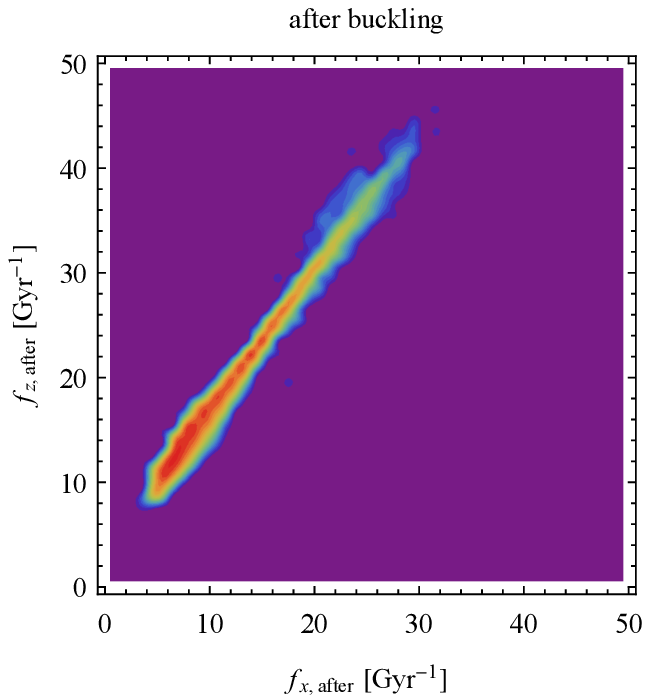}
\caption{Comparison of the properties of stellar orbits before (left column) and after (right column) buckling.
The first row of plots shows the distribution of values of the amplitude in the $z$ direction as a function of the
amplitude in $x$. The second row shows the distribution of stars in the plane of frequency ratio and amplitude of
oscillations in the $x$ direction. The third row shows the distribution of stars in the $f_z$-$f_x$ frequency plane.}
\label{beforeafter}
\end{figure*}

In Fig.~\ref{beforeafter} we compare the properties of the orbits before and after buckling (the left vs. right
column) in terms of their amplitudes and frequencies in the $x$ and $z$ directions. In the upper row we show the 2D
histogram of the amplitude in the $z$ direction as a function of the amplitude in $x$, $a_z (a_x)$. Before buckling
(left panel), the amplitudes in $z$ are much lower than those in $x$ since the bar is relatively flat. After buckling (right
panel), the distribution is very different, although the plot includes only those stellar orbits that actually buckled,
meaning that they all have $a_{z, {\rm \: after}} > a_{z, {\rm \: before}}$. Nevertheless, an interesting pattern is present: the
distribution is dominated by two branches: one up to $a_{x, {\rm \: after}} = 4.5$ kpc and a separate one at
larger $x$ amplitudes. The extent of the inner branch seems to correspond roughly to the size of the boxy/peanut shape
already present at this time (see the lower left panel of Fig.~\ref{surden}), while the outer branch corresponds
to the more distant part of the bar which is not yet part of the boxy/peanut shape but its $z$ amplitudes are
nevertheless significantly increased.

In the middle row of plots in Fig.~\ref{beforeafter} we show the distribution of stars in the plane of frequency ratio
$f_z/f_x$ and amplitude of oscillations in the $x$ direction, $a_x$. These maps contain similar information to the
histograms on the right of Fig.~\ref{histograms} but now with the dependence on $a_x$ added. We see that while before
buckling the distribution of frequency ratios is very wide, although somewhat narrower at smaller $a_x$, after buckling
it becomes very narrow with a very tight dependence on $a_x$. All stars follow a very tight correlation between
$f_z/f_x$ and $a_x$, with the exception of an additional small branch at $f_{z, {\rm \: after}}/f_{x, {\rm \: after}}
=2$ corresponding to banana-like orbits. Thus, when the dependence on $a_x$ is included, the range of occupied
frequency ratios is even narrower for the buckled bar than the lower right histogram of Fig.~\ref{histograms} would
suggest. We therefore confirm the result of \citet{Portail2015} who found that orbits of lower frequency ratios
$f_z/f_x$ occupy more inner parts of the boxy/peanut bar.

Further insight into the orbital structure of the bar before and after buckling can be obtained by plotting the
distribution of stars in the $f_z$-$f_x$ frequency plane at both times, as shown in the lower row of panels in
Fig.~\ref{beforeafter}. Before buckling, the distribution is much wider and the relation nonlinear, but after buckling,
the relation between the two frequencies becomes very tight and linear. However, there is no strict proportionality
between the two frequencies, that is the straight line does not cross the origin of the coordinate system. Instead, by
fitting a straight line to the data we find that it can be very accurately approximated by
\begin{equation}
        3 f_{z, {\rm \: after}} = 4 \ (f_{x, {\rm \: after}} + f_{\rm p})    \label{relation}
,\end{equation}
where $f_{\rm p}$ turns out to be equal to the pattern speed of the bar, often denoted by $\Omega_{\rm p}$, which in our case has
the value of 2.5 Gyr$^{-1}$. For $f_{x, {\rm \: after}}$ between 5 and 20 Gyr$^{-1}$ the
formula~(\ref{relation}) gives the values of $f_{z, {\rm \: after}}/f_{x, {\rm \: after}}$ between 2 and 1.5, exactly
as shown in the middle right panel of Fig.~\ref{beforeafter}. We note that when plotting all the frequencies $f_{z,
{\rm \: after}}$, $f_{y, {\rm \: after}}$ and $f_{x, {\rm \: after}}$ in a 3D plot rather than 2D, the frequencies lie
along two narrow branches corresponding to the two peaks of $f_{y, {\rm \: after}}/f_{x, {\rm \: after}}$ (see the lower
left panel of Fig.~\ref{histograms}), but when projected onto the $f_z$-$f_x$ frequency plane they both fall on the
same line.

The constant $f_{\rm p}$ obtained when fitting formula~(\ref{relation}) was simply a numerical value. However, it is
natural to expect that it is not just a random number but rather some characteristic frequency of the system, which we
indeed identify as the pattern speed of the bar. To make sure this is the case we made an independent calculation of
orbital frequencies in the traditional cylindrical coordinate system and estimated the angular, radial, and vertical
frequencies $\Omega$, $\kappa,$ and $\nu$ for the same selection of stars. We again found a very tight relation between
$\Omega$ and $\nu$. Since $f_x + f_{\rm p} = \Omega$ and $f_z = \nu$, the relation~(\ref{relation}) after buckling can
be written in an even simpler way as
\begin{equation}
        3 \nu = 4 \Omega.    \label{relation1}
\end{equation}

The results in terms of $\Omega$, $\kappa,$ and $\nu$ (not shown here) also confirm what was already demonstrated in
Fig.~\ref{histograms}, namely that most of the stars are on x1 orbits with $\kappa/(\Omega - \Omega_{\rm p}) = 2$
corresponding to the inner Lindblad resonance for the whole range of radii, both before and after buckling. Instead,
the vertical resonance, $\nu/(\Omega - \Omega_{\rm p}) = f_z/f_x = 2$ is populated only by a small fraction of stars
before and after buckling (see the middle panels of Fig.~\ref{beforeafter}). However, this resonance may have
contributed to initiating the buckling, as we discuss below.

\section{Discussion}

Using $N$-body simulations we have studied the buckling instability in a barred galaxy similar to the Milky Way. The
initial parameters of the model composed of an exponential disk embedded in a spherical dark matter halo were adjusted
so that the bar forms slowly. The buckling event, occurring after 4.5 Gyr from the start of the simulation, is however
quite violent and lasts only a fraction of a gigayear. We measured different properties of the buckling bar, including the
mean acceleration, velocity, and distortion in the vertical direction both as a function of radius
and in the face-on projection. We then studied the orbital structure of the bar before and after buckling determining
the apocenters and frequencies of the stellar orbits supporting the bar that undergo buckling. The key result of this
study is the discovery of a very tight relation between the frequency of stellar oscillations along the bar and in the
vertical direction. The relation explains the dependence of the frequency ratio on radius in boxy/peanut bars found
earlier in the literature. Still, the exact origin of the relation remains unclear.

We can however offer a tentative view on the nature of buckling instability using the results presented here. In
Fig.~\ref{azvzmeanz1} we show maps of the mean distortion of the stellar particles in the face-on view. In the
initial stage of buckling, at $t < 4.5$ Gyr (second row of plots in Fig.~\ref{azvzmeanz1}) there appears a
distortion with strong positive values at both ends of the bar, with maxima around $x = \pm 6$ kpc, and negative
distortion oriented approximately perpendicular to the bar. This configuration is stationary in the reference frame of
the bar for about 0.2 Gyr, that is it rotates together with the bar. Such a distortion can be produced by banana-like
orbits with typical apocenters around 6 kpc. We found that such orbits after buckling have frequencies consistent with
the vertical resonance $f_z/f_x = 2$. This suggests that the initial distortion of the bar is probably caused by
resonant trapping of x1 orbits that become banana-like orbits.

Such a distortion at time $t_0$ can be approximated in the circular form as $z_{\rm d} (R, \phi, t_0) = z_0(R) \cos m
\theta$ with $m=2$. As discussed by \citet{Binney2008} in the context of galactic warps (their section 6.6.1),
when neglecting the vertical velocity, the evolution of such a distortion at a given radius $R$ can be described as a
propagation of two kinematic bending waves with pattern speeds $\omega_{\rm p} = \Omega \pm \nu/2$ (a different symbol
$\omega_{\rm p}$ is used here to distinguish them from the pattern speed of the bar, $\Omega_{\rm p}$). After buckling,
the orbits in our bar obey $\nu = 4 \Omega/3,$ so the two pattern speeds are $\Omega/3$ (slow wave) and $5 \Omega/3$
(fast wave). The maximum distortion associated with the fast wave winds up rather quickly, as shown in the third row
of plots in Fig.~\ref{azvzmeanz1}. We may estimate the pitch angle $\alpha$ of the line of maximum height in the
usual way as $\cot \alpha = (5/3)\: R \: \Delta t \: |{\rm d} \Omega/ {\rm d} R|$ \citep{Binney2008}. For $\Delta t =
0.15$ Gyr (the time difference between $t=4.5$ and $t=4.65$ Gyr, the stages shown in the second and third rows of
Fig.~\ref{azvzmeanz1}) and $R=6$ kpc we find the pitch angle $\alpha \sim 30$ deg, in qualitative agreement with
the image shown in the third row of Fig.~\ref{azvzmeanz1}. On the other hand, the slow wave winds up much more
slowly in the inner part of the bar, and in the outer part, at $R=6-7$ kpc, it survives (lower row of
Fig.~\ref{azvzmeanz1}) because it corotates with the bar as in this region $\Omega/3 \approx \Omega_{\rm p}$.

The situation seems to be different in the case of the second buckling taking place between $t = 6.5$ and $t=8$ Gyr.
Although a similar pattern of distortion appears also in this case (see the upper row of panels in
Fig.~\ref{azvzmeanz2}), it occurs only in the outer part of the bar, at $R > 5$ kpc, but survives much longer.
Since the bar grows steadily in this time period (see Figs. \ref{a2profiles} and \ref{a2modestime}), it seems that as new
stars are captured by the bar they are at the same time or soon after trapped by the vertical resonance so that the
pattern is sustained. This picture is confirmed by the observation that the distortion moves toward outer radii in time
(Fig.~\ref{azvzmeanzprofiles}). In the end, around $t=8$ Gyr, this pattern also winds up leaving behind a much larger
boxy/peanut shape.

We conclude that buckling instability is probably essentially driven by the vertical resonance of the x1 stellar orbits
in the bar. The orbits trapped by the vertical resonance initiate the distortion of the bar which then evolves in the
form of kinematic bending waves. In the inner part of the bar the waves wind up forming a boxy/peanut shape, which in
the second buckling event is extended to larger radii. The details of this process certainly deserve further study.

\begin{acknowledgements}
Useful comments from the referee, Daniel Pfenniger, are gratefully acknowledged.
This work was supported in part by the Polish National Science Center under grant 2013/10/A/ST9/00023.
\end{acknowledgements}

\end{document}